\documentclass[11pt]{article}
\pdfoutput=1

\usepackage{booktabs}
\usepackage{jheppub}
\usepackage{siunitx}
\usepackage{slashed}

\newcommand\DLR[2]{\overset{\leftrightarrow}{D_{#1}^{#2}}}

\newcommand\re{\operatorname{Re}}
\newcommand{\lsim}{\raisebox{-0.13cm}{~\shortstack{$<$ \\[-0.07cm] $\sim$}}~} 
\newcommand{\gsim}{\raisebox{-0.13cm}{~\shortstack{$>$ \\[-0.07cm] $\sim$}}~} 
\newcommand\scalar{\varphi}
\newcommand\fermion{\chi}
\renewcommand\vector{\rho}
\newcommand\scalarTilde{\widetilde{\scalar}}

\newcommand\U{\rm U}
\newcommand\SU{\rm SU}
\newcommand\SO{\rm SO}

\title{A complete Effective Field Theory for Dark Matter}

\author[a]{Juan Carlos Criado}
\author[b,c]{Abdelhak Djouadi}
\author[b]{Manuel P\'erez-Victoria}
\author[b]{Jos\'e Santiago}

\affiliation[a]{Institute for Particle Physics Phenomenology, Department of Physics, Durham University,\\ 
Durham DH1 3LE, United Kingdom.}

\affiliation[b]{Centro Andaluz de Fisica de Particulas Elementales (CAFPE) \\ and \\ Departamento de F\'isica Te\'orica y del Cosmos, \\ Universidad de Granada, E--18071 Granada, Spain.}

\affiliation[c]{Laboratory of High Energy and Computational Physics, NICPB, R\"avala pst. 10,\\ 10143 Tallinn, Estonia.}

\abstract{
We present an effective field theory describing the relevant interactions of the Standard Model with an electrically neutral particle that can account for the dark matter in the Universe. The possible mediators of these interactions are assumed to be heavy. The dark matter candidates that we consider have spin 0, 1/2 or 1, belong to an electroweak multiplet with arbitrary isospin and hypercharge and their stability at cosmological scales is guaranteed by imposing a $\mathbb{Z}_2$ symmetry. We present the most general framework for describing the interaction of the dark matter with standard particles, and construct a general non-redundant basis of the gauge-invariant operators up to dimension six. The basis includes multiplets with non-vanishing hypercharge, which can also be viable DM candidates. We give two examples illustrating the phenomenological use of such a general effective framework. First, we consider the case of a scalar singlet, provide convenient semi-analytical expressions for the relevant dark matter observables, use present experimental data to set constraints on the Wilson coefficients of the operators, and show how the interplay of different operators can open new allowed windows in the parameter space of the model. Then we study the case of a lepton isodoublet, which involves co-annihilation processes, and we discuss the impact of the operators on the particle mass splitting and direct detection cross sections. These examples highlight the importance of the contribution of the various non-renormalizable operators, which can even dominate over the gauge interactions in certain cases.
}

\begin{document}

\maketitle
\flushbottom

\section{Introduction}

The existence of dark matter (DM) in the universe is strongly suggested by several astrophysical and cosmological measurements but its very nature remains enigmatic. Particle physics proposes a plausible and effective solution to this problem in terms of an electrically neutral and weakly interacting massive particle that is stable at cosmological scales \cite{Bertone:2004pz,Drees:2012ji}. DM particles are predicted by many extensions of the Standard Model (SM), including the well motivated ones that address other important theoretical or experimental issues of the model such as supersymmetric~\cite{Ellis:1983ew,Goldberg:1983nd} or extra-dimensional models~\cite{Servant:2002aq,Cheng:2002ej,Agashe:2007jb,Panico:2008bx}.\footnote{In these extensions the dark matter candidate is the lightest particle of the new sector, which can be made stable by virtue of a discrete symmetry. This symmetry also helps in relaxing direct and indirect collider limits.} In fact, basically any extension of the SM with additional neutral particles can accommodate a DM state, provided that a discrete symmetry is imposed to protect it from decaying into SM particles. These DM particles are turning out to be the new holy Grail of contemporary physics, actively searched in many astroparticle \cite{Aghanim:2018eyx,Aprile:2018dbl,Aprile:2019xxb,Aprile:2019dbj,Agnes:2018ves,Aalbers:2016jon} and collider \cite{Aad:2015pla,Khachatryan:2016whc,Aaboud:2019rtt,Sirunyan:2018owy,Khachatryan:2014rra} experiments. 

Because of the large proliferation of DM candidates, it has become customary and quite useful to consider effective field theory (EFT) approaches, which allow to study in a model-independent manner the phenomenology of these particles. It is typically assumed that the new state is either a scalar, a vector or a fermion, although higher spins have been also considered, see refs.~\cite{Criado:2020jkp,Falkowski:2020fsu} for recent accounts. Among the simplest and most economical of these EFTs are the Higgs-portal models, in which a single DM particle is introduced in addition to the SM particles; it interacts in pairs only with the Higgs sector of the theory, which is assumed to be minimal \cite{Djouadi:2005gi} and, hence, involves only the unique Higgs boson observed at the LHC \cite{Aad:2012tfa,Chatrchyan:2012ufa}; see refs.~\cite{Arcadi:2019lka,Arcadi:2021mag} for  recent reviews. There are also other possibilities for these simple EFTs and, for instance, $Z$--portal \cite{Cotta:2012nj,Arcadi:2014lta,Balazs:2017ple} or neutrino--portal \cite{Gonzalez-Macias:2016vxy,Escudero:2016tzx,Batell:2017cmf} scenarios have been also extensively discussed. These models are rather predictive as all DM observables can be specified by only a few basic new parameters, for example, the mass of the DM particle and its coupling with the mediator particle.  

However, it might well be that the relevant new physics extension has also additional new particles, heavy scalars, vectors or fermions  that could accompany the DM and/or act as DM mediators. Actually, the simple portal models, with the mediator being exclusively a SM particle, are highly constrained by present data. In this case, another possible approach which also introduces only a handful of free parameters has been put forward:  simplified models \cite{Alves:2011wf,Abdallah:2014hon} in which an effective Lagrangian is introduced that explicitly includes the mediator particle and its interactions with the SM and DM states. For instance, a model with a singlet DM fermion and a singlet scalar mediator has been studied in ref.~\cite{Baek:2015lna} and the importance of imposing the full SM gauge symmetry is emphasized; other examples of simplified models can be found in refs.~\cite{Bell:2015rdw,Cyr-Racine:2015ihg,Goncalves:2016iyg,DeSimone:2016fbz,Alanne:2017oqj,Alanne:2020xcb}. This hybrid approach has many advantages, in particular in the context of collider physics when the mass of the DM and its possible companions are comparable to the collider energy, but in practice introduces, unlike the simple EFT approach with only DM fields, some additional model dependence that makes it rather close to the concrete realizations that it is meant to simplify. We will thus not discuss this approach further here. 

Instead, in this paper, we follow a genuine EFT approach, including all the operators allowed by the symmetries. We assume that all possible mediators of the DM interactions are heavy and have been integrated out. The degrees of freedom in the EFT are only the SM particles and the DM particle, together with its possible gauge partners, which will be nearly-degenerate in mass. Indeed, whenever the DM field has non-vanishing isospin, it must appear together with other fields, forming complete $\SU(2)_L$ multiplets. In our EFT we consider a DM particle of spin 0, $\frac12$ or 1, with well-defined gauge quantum numbers. That is, the DM field appears in the neutral component of an $\SU(2)_L \times \U(1)_Y$ multiplet.  For these three different spin assignments and for arbitrary isospin and hypercharge, we consider the most general EFT that describes this scenario, with the additional assumption of a $Z_2$ symmetry to stabilize DM. This is nothing but an extension of the SMEFT~\cite{Brivio:2017vri} with new degrees of freedom: an additional scalar, vector or fermionic multiplet containing the DM state.\footnote{Due to the $Z_2$ symmetry, the EFT in this paper is complementary to the one considered in ref.~\cite{deBlas:2017xtg}, in which the extra fields in the SM extension were required to have linear interactions with SM operators.} We perform a systematic classification of the gauge-invariant operators that appear in this extended EFT, and construct a general basis for all relevant operators up to dimension six. These operators capture the effects of the mediators or additional heavy particles, and also of possible non-perturbative UV completions~\cite{Bruggisser:2016ixa}. The Wilson coefficients of these operators, to be treated as free parameters, can be constrained using present data on the cosmological relic density, direct and indirect DM detection in astroparticle physics experiments and in missing energy searches or invisible Higgs boson decays at high-energy colliders like the LHC.

DM multiplets with non-vanishing hypercharge are often discarded because their scattering off nucleons have large cross-sections mediated by a $Z$ boson, in conflict with direct-detection bounds. However, it is known that the dominant contribution to the cross-section is avoided if the DM particle is described by a Majorana spinor~\cite{Kayser:1981nw}. This is the case of fermionic DM particles with non-vanishing hypercharge whenever the two components of the Dirac field are non-degenerate. The mass splitting can be produced by mixing with an additional Majorana fermion. More relevant to our EFT setup with one multiplet, it has also been shown in the case of a fermion doublet that it can be induced by non-renormalizable operators~\cite{Dedes:2016odh}. Here, we argue that this mechanism is very general and works for arbitrary spin whenever there is some breaking of a global U(1) symmetry acting on the DM multiplets. We distinguish the operators in our EFT that break this symmetry and observe that, to dimension 6, they require fields of isospin 1/2.

There have been many analyses in the past that have considered EFTs whose field content were that of the SM extended with DM states.  In the low-energy regime, non-relativistic EFTs have been discussed, for example in refs.~\cite{Fan:2010gt, Fitzpatrick:2012ib,Fitzpatrick:2012ix, Bellazzini:2013foa, DelNobile:2013sia, Catena:2014uqa,Catena:2014epa,Ovanesyan:2014fwa,Schneck:2015eqa,Catena:2019hzw,Nobile:2021qsn}, but more focus has been put on the relativistic case. For instance, a complete set of operators of dimension $\leq 6$ (written in the broken phase of the electroweak sector) for an EFT containing the SM fields together with a complex scalar DM field in a singlet, doublet or triplet representation is given in ref.~\cite{DelNobile:2011uf}. In ref.~\cite{DeSimone:2013gj}, an EFT consisting of the SM particles together with Majorana or real scalar DM has been considered and the operators of dimension $\leq 8$ have been classified in the case where the DM particle is assumed to be a singlet under the SM gauge group and coupling only to fermions. In ref.~\cite{Duch:2014xda}, a basis of operators of dimension $\leq 6$ for an EFT with the SM and DM particles has been given, with the non-SM fields being a light right-handed neutrino as well as a singlet scalar, fermion and vector DM fields which were made stable by invoking a $\mathbb{Z}_2$ symmetry under which the SM fields are uncharged and the DM fields are charged.
A basis of dimension-6 operators describing interactions of a singlet-like Majorana DM fermion with SM particles, and its applications for collider and astroparticle searches, has been introduced in Refs.~\cite{Matsumoto:2016hbs,Matsumoto:2014rxa,Han:2017qkr}. Closer to our scope in this paper, a minimal basis of operators of dimension 6 or less describing the interaction of a singlet scalar, a Dirac fermion and a vector DM with quarks and gluons only has been given in ref.~\cite{Belyaev:2018pqr}. Finally, in ref.~\cite{Brod:2017bsw}, a basis of effective operators up to dimension 7 has been given in the case of scalar and fermionic DM embedded in a general SM multiplet.

Subsets of effective operators relevant for different applications have been considered in refs.~\cite{Harnik:2008uu, Kopp:2009qt, Goodman:2010qn, Cheung:2012gi,Buckley:2013jwa, Crivellin:2014gpa, Crivellin:2015wva, DeSimone:2016fbz, Fedderke:2014wda, Hisano:2015bma,Bhattacharya:2021edh}.
In particular, DM interactions with quarks and gluons and their impact on direct detection as well as searches at colliders have been discussed in refs.~\cite{Fox:2011pm,Goodman:2010yf,Goodman:2010ku} in the case of singlet Dirac and/or Majorana DM and in ref.~\cite{Goodman:2010ku} in the case of scalar DM. The specific case of fermion DM couplings to photons was considered in ref.~\cite{Arina:2020mxo}. An EFT for fermionic and scalar DM interactions with quarks, gluons and photons at low energy scales has been presented in ref.~\cite{Bishara:2016hek}.
In ref.~\cite{Crivellin:2014qxa}, loop effects in an EFT of the SM extended with a singlet Dirac fermion DM particle have been studied. The matching at the electroweak scale between the DM EFTs with and without the standard fermion, gauge and Higgs bosons has been performed in ref.~\cite{Hill:2014yka}.
Finally, the important aspect of co-annihilation has been also discussed and, for instance, an EFT for the DM in this context was constructed in ref.~\cite{Bell:2013wua} (while a systematic classification of simplified models for co-annihilation has been given in ref.~\cite{Baker:2015qna}).

In this paper, we complete the previous analyses by presenting the most general framework for the description of the interactions between one multiplet that contains the DM particle and SM particles, which we assume to be the only relevant degrees of freedom at sufficiently low energies. The EFT for the DM-SM system in our setting is obtained by extending the SMEFT field content and symmetries, which besides the Lorentz group and the SM gauge invariance include a  discrete $\mathbb{Z}_2$ symmetry in order to stabilize the DM particle. As mentioned before, we study the case of spin-0, $\frac12$ and 1 DM, not constrained to be a SM singlet. 
In particular, we extend in several ways on the work of ref.~\cite{Brod:2017bsw}, which to the best of our knowledge contained the most complete list of operators for generic DM available to this date. First, we discuss the case of vector DM multiplets in addition to the scalar and fermionic cases. Second, we do not assume the dark $\U(1)_D$ symmetry considered in ref.~\cite{Brod:2017bsw}, which allows for extra operators for multiplets with hypercharge $Y = 1/2$.\footnote{We thank the authors of ref.~\cite{Brod:2017bsw} for a discussion on this issue.} Third, our non-redundant basis for scalar multiplets with arbitrary hypercharge $Y\neq 1/2$ includes two  operators that were not considered in ref.~\cite{Brod:2017bsw}. Finally, we  present some interesting phenomenological implications of our approach and, in particular, we address the important consequences of the additional operators for multiplets with hypercharge $Y= 1/2$. As we explain below, these operators are required to produce the mass splitting that helps in avoiding direct-detection limits. On the other hand, we stop at dimension 6 and do not consider dimension-7 operators, which have been included in ref. ~\cite{Brod:2017bsw}.

The phenomenology of singlet and non-singlet DM particles is quite different and we give an example of each possibility to illustrate the use of our effective theory and to show the relevance of the non-renormalizable operators. We first consider the case of a singlet scalar DM state, where we give semi-analytical expressions for all the relevant DM observables, which are rather simple and convenient for practical use. Indeed, these expressions allow a more efficient exploration of the effect of the various operators and, as an example, we use them to study the possible cancellation of the contributions of two different operators, which allow to ease the stringent experimental constraint from direct DM detection. In a subsequent step, we illustrate the important impact of co-annihilation by considering  the case of an iso-doublet of heavy leptons and discuss the simultaneous effect of the various dimension-5 operators on the mass splitting between the neutral DM state and the other charged and neutral leptons, as well as on the annihilation and co-annihilation cross sections into SM particles. 

The paper is organized as follows. In the next section, we discuss the derivation of the complete basis of operators of dimension $\leq 6$ which involve two scalar, fermionic or vector DM multiplets and SM fermions, gauge and Higgs bosons. In sections~\ref{sec:scalar} and~\ref{sec:lepton-example}, we  use our EFT to analyze the phenomenology of a scalar singlet and of a fermion doublet, respectively. A short conclusion is given in section~\ref{sec:conclusions}.
In Appendix~\ref{app:bases},  we present our complete basis of operators up to dimension 6 and in Appendix~\ref{app:operators} we comment on the basis of operators of the form $\scalar^\dagger \scalar \, \phi^\dagger \phi D^2$.


\section{Effective field theory for a generic DM multiplet}
\label{sec:eft}

\subsection{Symmetries and field content}

Our aim in this study is to set up a general framework for the description of the interactions between the DM and the SM particles in the case in which the possible mediators of the interactions are heavier than these particles and than the energies which are expected to be probed. This DM-EFT can be obtained by extending the SMEFT (the effective field theory constructed with the SM fields only) with an extra field content, a multiplet that we will generically denote $\mathcal{X}$, that involves the DM particle, $\mathcal{X}_0$, and it possible companions. The SM gauge group $G_{\text{SM}} = \SU(3)_C \times \SU(2)_L \times \U(1)_Y$ must be contained in the symmetry group of this EFT, so we shall impose the SM gauge symmetry without losing generality. Our only assumption in this regard is that $G_{\text{SM}}$ is linearly realized (as it is the case in the SMEFT) but, of course, we shall also impose Lorentz symmetry. Thus, the extra fields of the DM-EFT can be organized into irreducible representations of $G_{\text{SM}}$ and the Lorentz group. As any EFT, our theory is valid only below a cut-off $\Lambda$, which should be larger than the electroweak scale scale $v \approx 246$ GeV as well as the DM mass $M_{\mathcal{X}}$. When introducing the various operators that describe the interactions in this EFT, we implement the usual power counting assigning a typical value of $ \Lambda^{4 - \Delta}$ to the Wilson coefficient of an operator of dimension $\Delta$.

In order to work with a manageable theory some restrictions on the DM sector need to be imposed. In this paper we make the following assumptions:
\begin{enumerate}
\item  In order to stabilize the DM particle, we impose a discrete  $\mathbb{Z}_2$ symmetry. 

\item The field content of the theory is given by the SM one, including the Higgs doublet $\phi$, and a {\em single} extra multiplet $\mathcal{X}$ that belongs to an irreducible representation of $G_{\text{SM}}$. Under the Lorentz group, $\mathcal{X}$ transforms either as a scalar, a spinor or a vector. All SM fields are even under $\mathbb{Z}_2$, while $\mathcal{X}$ is odd. 
\end{enumerate}

Due to the restriction to one extra multiplet, our EFT can be regarded as a minimal extension of the SMEFT, much as the minimal dark matter scenario~\cite{Cirelli:2005uq,Cirelli:2009uv,Bottaro:2021srh} is a minimal extension of the SM. We make this assumption for simplicity but, in fact, the same EFT works in the case of several flavors of the same multiplet and also when there is a separate $\mathbb{Z}_2$ for each type of multiplet; only an additional labelling of the different fields and couplings would be required. The general case with different types of multiplets would instead involve extra operators, which we do not write here. 

If the cutoff $\Lambda$ is large, the assumption of a $\mathbb{Z}_2$ symmetry is not necessary for multiplets in sufficiently large representations of $G_{\text{SM}}$, as the gauge symmetry then forbids operators linear in  $\mathcal{X}$ below a given dimension\footnote{For instance, note that no renormalizable operator linear in $\mathcal{X}$ is allowed for representations not included in the list which is given in e.g. ref.~\cite{deBlas:2017xtg}.}. The DM particle is then not absolutely stable and could decay, but  the corresponding suppression of the decay width could make the lifetime longer than the age of the Universe, as emphasized in ref.~\cite{Cirelli:2005uq}. However, we are mostly interested in the case in which $\Lambda$ is not much larger than the TeV scale, so the suppression will not be sufficient except for extremely large representations.  Finally, the restriction to spin up to unity is also made for simplicity; DM with higher spins has been recently considered in refs.~\cite{Criado:2020jkp,Falkowski:2020fsu}.

In order to include a good DM candidate, $\mathcal{X}$ must be a color singlet and must contain an electrically-neutral component $\mathcal{X}_0$. This component is the field associated to the DM particle(s). The existence of $\mathcal{X}_0$ restricts the possible electroweak quantum numbers of $\mathcal{X}$. Let $T$ be its isospin and $Y$ its hypercharge. Then, an electrically neutral component is present if and only if the difference $T - |Y|$ is a natural number. We assume that $Y$ is non-negative without loss of generality: from a multiplet $\mathcal{X}$, one can obtain a new one $\widetilde{\mathcal{X}}$ in the same representation of $\SU(2)_L$ and with opposite hypercharge as
\begin{equation}
    \label{eq:SU2-conjugation}
    \widetilde{\mathcal{X}} = \mathcal{E} \hat{\mathcal{X}},
\end{equation}
where $\mathcal{E}$ is the anti-diagonal matrix with entries $(1, -1, 1, -1, \ldots)$ and $\hat{\mathcal{X}}$ is the conjugate of the field $\mathcal{X}$ ($\hat{\scalar}=\scalar^*$, $\hat{\fermion}=\gamma_0 C \fermion^*$ and $\hat{V}_\mu=V_\mu^*$ for scalar, fermion and vector fields, respectively).\footnote{This is related to charge conjugation $\mathcal{C}$ by $\mathcal{C} \mathcal{X} \mathcal{C}^{-1} = \eta^C_{\mathcal{X}} \hat{\mathcal{X}}$, $\eta^C_{\mathcal{X}} = \pm 1$.} We thus have an infinite but discrete set of possibilities for the DM multiplet $\mathcal{X}$, given by
\begin{equation}
  T \in \{0, \; 1/2, \;  1, \; \ldots\}, \qquad Y \in \{T, \; T - 1, \; \ldots, \; T - \lfloor T\rfloor\},
\end{equation}
where $\lfloor T\rfloor$ denotes the integer part of the isospin $T$.
The minimal field content of multiplets with vanishing hypercharge and integer isospin corresponds to irreducible representations carried by real scalars, Majorana spinors or real vectors.

We note that multiplets with large $T$ values will accelerate the running of the $\SU(2)$ gauge coupling $g$ (and the $\U(1)$ coupling $g^\prime$ if the hypercharge $Y$ is also large). We do not impose any restriction in this sense because $\Lambda$ is arbitrary and the scale at which $g$ becomes non-perturbative will typically be beyond the regime of validity of the EFT.

Our approach covers a large number of the models, both complete in the ultraviolet regime and constructions in the effective theory approach,  that have been proposed in the literature and involving a weakly interacting massive DM particle.  Some of these scenarios have been mentioned before and, for instance, models in which the DM interacts with the SM particles through the Higgs portal have been reviewed recently in ref.~\cite{Arcadi:2019lka} where a rather exhaustive list of references may be found.

For most phenomenological purposes, the only relevant operators are those that contain the DM field quadratically. The kinetic and mass terms read
\begin{equation}
    \mathcal{L}_0 =\;
  \eta_\scalar \big[\big(D_\mu \scalar\big)^\dagger D^\mu \scalar
  - M_\scalar^2 \scalar^\dagger \scalar \big],
\end{equation}
for a scalar multiplet, denoted by $\scalar$;
\begin{equation}
    \mathcal{L}_0 =\;
  \eta_\fermion \big[\bar{\fermion} i \slashed{D} \fermion
  - M_\fermion \bar{\fermion} \fermion \big],
  \label{eq:quad-lag-fermion}
\end{equation}
for a fermion multiplet, denoted by $\fermion$; and
\begin{equation}
   \mathcal{L}_0 =\;
  \eta_\vector \big[
  \big(D_\mu \vector_\nu\big)^\dagger D^\nu \vector_\mu
  - \big(D_\mu \vector_\nu\big)^\dagger D^\mu \vector_\nu
  + M_\vector^2 \vector_\mu^\dagger \vector^\mu \big],
\end{equation}
for a vector multiplet, denoted by $\vector$. The derivatives $D_\mu$ are covariant with respect to $G_{\text{SM}}$ and $\eta_\mathcal{X}=1,\frac{1}{2}$ for complex and real representations of the multiplet $\mathcal{X}$, respectively. Note that the spin-one extra particles are described in our effective treatment by a Proca vector field. This vector field can represent an extra gauge field of an extended gauge group;\footnote{This is actually necessary for perturbativity if the cutoff is not close to the mass of the spin 1 particle.} in our minimal setup, this comes along with the assumption that the Higgs fields associated to the spontaneous breaking of the additional gauge invariance are heavier than the scale $\Lambda$, and have been integrated out. In order to allow for non-perturbative UV completions, we do not impose here the restrictions on the Wilson coefficients that would arise from the extra gauge invariance.

A list of all the operators with dimension $D\leq 6$ for any electroweak multiplet $\mathcal{X}$ is given in appendix~\ref{app:bases}. This is the main result of this paper. The effective Lagrangian can thus be written as
\begin{equation}
    \mathcal{L}_{\text{DMEFT}}
    =
    \mathcal{L}_{\text{SMEFT}}
    + \mathcal{L}_0
    + \sum_i c^{\text{H}}_i \mathbf{O}^{\text{H}}_i
    + \sum_i \left(c^{\text{NH}}_i \mathbf{O}^{\text{NH}}_i + \text{h.c.}\right),
\end{equation}
where $\mathcal{L}_{\text{SMEFT}}$ is the SMEFT Lagrangian; the $\mathbf{O}^{\text{H}}_i$ and $\mathbf{O}^{\text{NH}}_i$ operators are the hermitian and non-hermitian operators in appendix~\ref{app:bases} respectively. Also, in the equation above, the $c^{\text{H}}_i$ coefficients are real, while the $c^{\text{NH}}_i$ are complex.

Not all structures are allowed for all choices of isospin and hypercharge. The restrictions arise because the quantum numbers of the product of two $\mathcal{X}$ must match the quantum numbers of the combination of SM fields appearing in a given operator. For instance, when $D\leq 6$, whenever the latter has non-vanishing hypercharge, $\mathcal{X}$ must have $Y=1/2$ for the operator to be gauge invariant. These tables comprise a non-redundant basis of operators: gauge-invariant operators not written here can be expressed as linear combinations of these by the use of algebraic identities, integration by parts and field redefinitions. 

\subsection{Suppressing gauge couplings \label{sec:Z-couplings}}
The dark matter fields in multiplets with $Y=0$ obviously have no coupling to the $Z$ boson. On the other hand, multiplets with $Y\neq 0$ are usually avoided because in general gauge interactions, and in particular the coupling to the $Z$ boson, give too large a contribution to the direct detection cross section, well above the current experimental limits. However, as we discuss in this section, there is a quite generic scenario in which the vector coupling to the $Z$ boson, and therefore its contribution to spin-independent direct detection, vanishes. This leaves only gauge contributions to spin-dependent direct detection processes that are much less constrained experimentally.

The electrically neutral component of multiples with $Y \neq 0$ is a complex field $\mathcal{X}_0$, which transforms trivially under $\U(1)_Q$ but non-trivially under $\U(1)_Y$. Hence, it has gauge couplings to the Z boson field with strength $e Y/(s_W c_W)$. This complex field is made of two self-conjugate components: $\mathcal{X}_0 = A + i B$, 
with $\hat{A}=A$, $\hat{B}=B$.
Consider now the mass term of the fields $A$ and $B$. It is a hermitian form over self-conjugate fields, so it is equivalent to a quadratic form and the 2x2 mass matrix $\mathcal{M}$ is real and symmetric. This is diagonalized by a real orthogonal transformation and its eigenfields $N_1$ and $N_2$ are also self-conjugate. If $\mathcal{M}$ has two degenerate eigenvalues, then $A$ and $B$ are mass eigenfields as well, and the full complex $\mathcal{X}_0$ constitutes the DM candidate. On the other hand, if the eigenvalues are non-degenerate, then only the particle associated to the lighter mass eigenfield, $N_1$, survives and forms the DM today. In this case, only $N_1$ is relevant as an initial state in DM-nucleon interactions.
Furthermore, if the mass separation is larger than the typical kinetic energy of a DM particle near the Earth, only $N_1$ will appear as a final state. 

Let us consider the non-degenerate case. We are interested here in the trilinear interaction of two $N_1$ fields and one Z boson field, which will be of the form $j^{N_1}_\mu Z^\mu$. The current $j^{N_1}_\mu$ is necessarily invariant under $\mathcal{C}$, since it is bilinear in $N_1$ and this field is self-conjugate. Hence, since $A_\mu$ is $\mathcal{C}$ odd, these trilinear interactions cannot conserve $\mathcal{C}$. If the interaction preserves $\mathcal{C}P$, it must be parity-odd. That is, $j^{N_1}_\mu$ must be a pseudovector current. Vector currents that couple to the Z boson preserving $\mathcal{C}P$ (as the ones from minimal coupling) cannot be built with $N_1$ alone. These are the currents that enter the spin-independent nuclear-DM cross-sections mediated by a $Z$, which thus vanish for non-degenerate $N_i$. In this scenario, small contributions from non-renormalizable operators can dominate direct detection.

The relevant question is then: when are the two mass eigenfields non-degenerate? Consider a global $\U(1)_X$ symmetry transformation acting only on the DM multiplet $\mathcal{X}$. This acts as an $\SO(2)$ transformation on the two-dimensional real vector space generated by $A$ and $B$. In fact, $A$ and $B$ span the vector space of a two-dimensional irreducible representation of this $\SO(2)$, since no one-dimensional subspace is left invariant. Indeed, a generic $\SO(2)$ transformation transforms any vector into a linearly independent one. The vector space of the irreducible representation is thus the complete space generated by $A$ and $B$. If the global $\U(1)_X$ is a symmetry of the action, then the mass matrix $\mathcal{M}$ will commute with the $\SO(2)$ transformations and all the linear combinations of $A$ and $B$ will be mass eigenfields with the same eigenvalue.\footnote{An equivalent way to put this is that a $2 \times 2$ matrix left invariant by arbitrary rotations must be a multiple of the identity.}  So, if $\U(1)_X$ is preserved, we are in the degenerate case. 

Conversely, any breaking of $\U(1)_X$ will generically produce a splitting of the eigenmasses. This is the case, for instance, of SUSY DM models in which the Higgsino mixes with a Majorana gaugino, which breaks the symmetry. In our effective treatment with only one DM multiplet, the breaking of $\U(1)_X$ can only occur in operators in which the combination of DM fields has non-vanishing hypercharge. The reason is that $\U(1)_X$ acts as a global $\U(1)_Y$ on the DM fields. For operators that are bilinear in $\mathcal{X}$, the only possible combinations with non-vanishing hypercharge are products $\mathcal{X} \mathcal{X}$ or $\hat{\mathcal{X}}\hat{\mathcal{X}}$, with hypercharge $\pm 2Y$. From gauge invariance, they must multiply a SM operator with hypercharge $\mp 2Y$. Hence, the $Y$ assignments for which $\U(1)_X$-breaking operators of a given dimension exist are  restricted by the hypercharge of the possible SM operators that can enter those operators. To dimension 6, the only possibility is $Y\!=\!1/2$. From the point of view of direct detection, these multiplets can behave similarly to the ones with $Y\!=\!0$.

This symmetry argument applies also at the quantum level, since a linear global $\U(1)_X$ preserved by the classical action is non-anomalous--- the left-handed and right-handed components of Dirac fermions transform in the same way---and will also be a symmetry of the quantum effective action, and in particular of the quadratic terms in the effective potential. 

As we have emphasized, this mechanism works for bosons as well as for fermions. Nevertheless, spin-independent contributions to nuclear-DM scattering at low energies are suppressed in any case, (in particular for $Y\neq 1/2$) when $\mathcal{X}$ is a boson, as the couplings of $\mathcal{X}_0$ to the $Z$ boson necessarily involve derivatives.


\section{Scalar singlet dark matter}
\label{sec:scalar}

In order to study the effects of each effective operator, one needs to fix a concrete multiplet. In this section, we will use a real scalar singlet $\scalar$ \cite{Kanemura:2010sh,Djouadi:2011aa,Djouadi:2012zc,Silveira:1985rk,McDonald:1993ex,McDonald:1993ex,Burgess:2000yq,Barger:2008jx,Andreas:2010dz,Baek:2014jga,Athron:2017kgt,Gross:2017dan} as an example of how to use the general framework introduced in the previous section. A similar analysis can also be performed for fermion and vector singlets. The case of non-singlet multiplets will be illustrated in section~\ref{sec:lepton-example}. For the scalar singlet, there are 12 independent allowed operators, shown in the first column of table~\ref{tab:operators-singlet}.

\begin{table}
\renewcommand{\arraystretch}{1.20}
  \centering
  \begin{tabular}{cccccccccccc}
    \toprule
    Operators
    & Contraints \\
    \midrule
    $\mathbf{O}_{\phi 1}$, $\mathbf{O}_{\phi 4}$
    & $H \to \text{inv}$, direct detection \\
    $\mathbf{O}_{\phi \square 1}$ & $H \to \text{inv}$ \\
    \midrule
    $(\mathbf{O}_{e\phi 1})_{33}$ & -- \\
    $(\mathbf{O}_{d\phi 1})_{33}$, $(\mathbf{O}_{u\phi 1})_{33}$
    & direct detection \\
    \midrule
    $\mathbf{O}_B$, $\mathbf{O}_{W1}$,
    $\mathbf{O}_{\widetilde{B}}$, $\mathbf{O}_{\widetilde{W}1}$
    & -- \\
    $\mathbf{O}_G$, $\mathbf{O}_{\widetilde{G}}$
    & monojet \\
    \bottomrule
  \end{tabular}
  \caption{Allowed operators for a real scalar singlet DM candidate and sources
    of constraints over their coefficients. All operators are relevant for setting the correct DM abundance. They all have dimension 6, except for $\mathbf{O}_{\phi 1}$, which has dimension 4.}
  \label{tab:operators-singlet}
\end{table}

The relevant experimental constraints over the coefficients of the operators for the scalar singlet $\scalar$ are the measured value of the DM relic abundance and the upper limits on: the nucleon-DM cross section from direct detection experiments; the DM annihilation cross section from indirect detection; the Higgs branching ratio into invisible decay products and the monojet cross-section from the LHC. For the relic abundance, we take Planck collaboration's~\cite{Aghanim:2018eyx} value of $0.120$. For the nucleon cross section we use the bounds from DarkSide-50~\cite{Agnes:2018ves} and Xenon1T~\cite{Aprile:2018dbl} experiments, for masses below and above $\SI{6}{GeV}$, respectively. For the DM annihilation cross section into $b\bar{b}$ and $W^+W^-$ final states, we apply the bound obtained in ref.~\cite{Boddy:2019qak} using the data on the gamma ray flux from dwarf galaxies. For the collider limits, we use the LHC results from ATLAS and CMS on the invisible Higgs branching ratios given in refs.~\cite{Aaboud:2019rtt,Sirunyan:2018owy} and on the missing energy events given by CMS in ref.~\cite{Khachatryan:2014rra}. 

\begin{figure}
  \centering
  \includegraphics[width=0.98\linewidth]{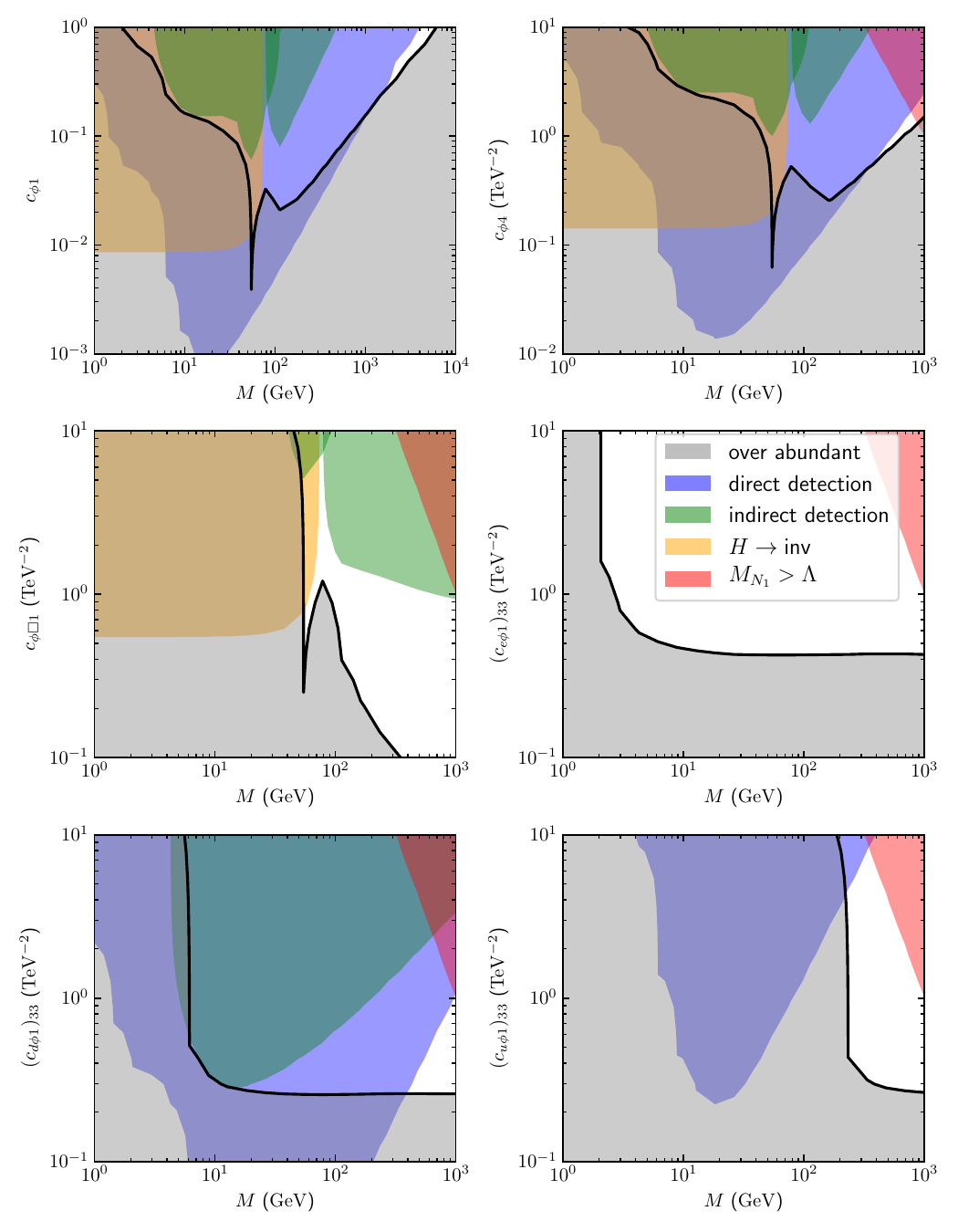}
  \caption{Constraints on the coefficients of the operators coupling the DM particle to the Higgs boson and/or SM fermions, as a function of the DM particle mass. The points over the solid black line give the correct DM abundance. The gray region gives an overabundance of DM. The orange region is excluded by the bound on $H \to \text{invisible}$ from ATLAS and CMS. The blue region is excluded by the XENON1T and the DarkSide-50 experiments. The region excluded by the limit on the annihilation cross section set by Planck is shown in green. The EFT is no longer valid in the red region.}
  \label{fig:limits-1}
\end{figure}

\begin{figure}
  \centering
  \includegraphics[width=0.98\linewidth]{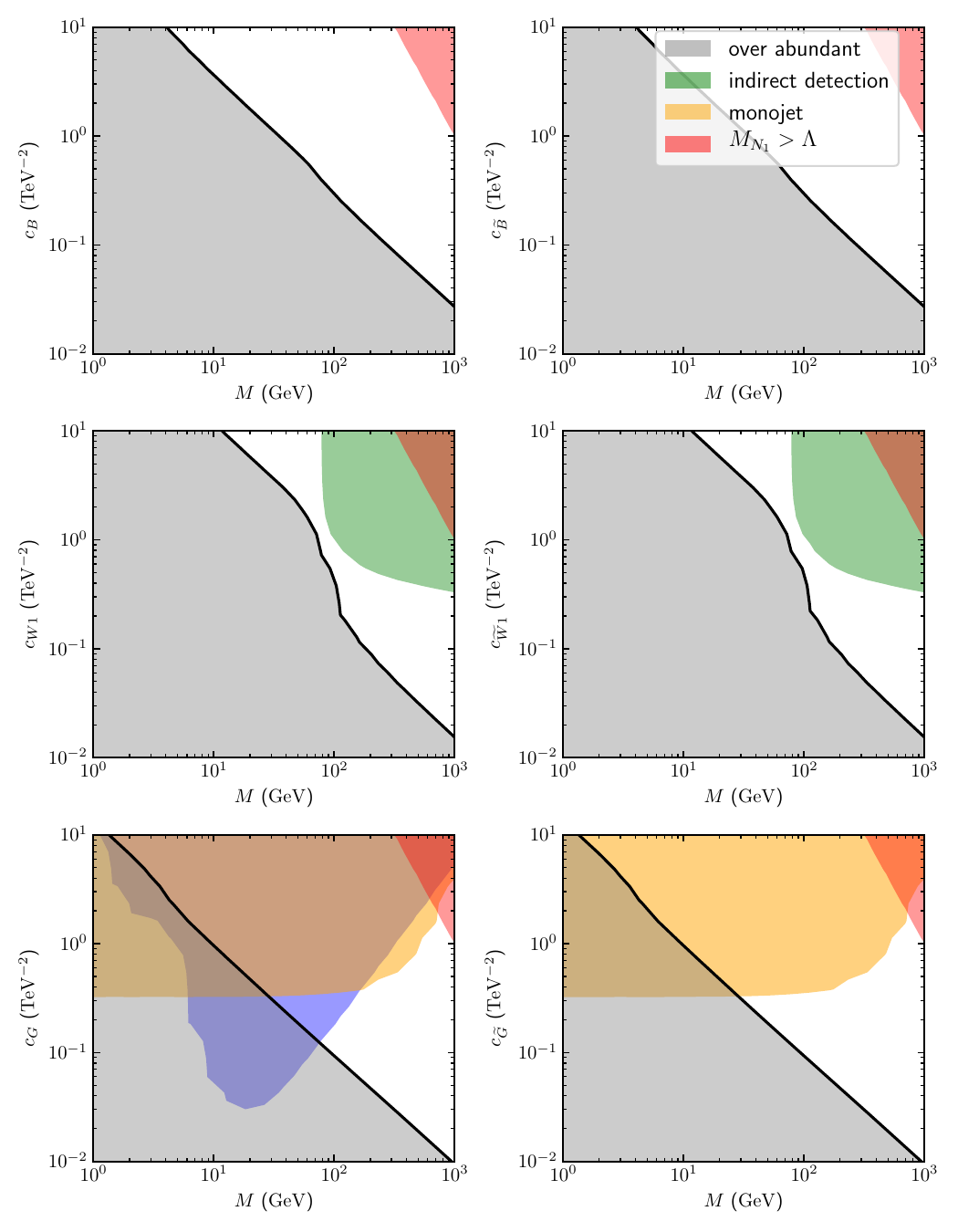}
  \caption{Constraints on the coefficients of the operators coupling the scalar DM particle to gauge bosons, as a function of the DM particle mass. The points over the solid black line give the correct DM relic abundance. The gray region gives an overabundance of DM. The orange region is constrained by the analysis of monojet events performed by the CMS collaboration. The region excluded by the limit on the annihilation cross section set by Planck is shown in green. The EFT is no longer valid in the red region.}
  \label{fig:limits-2}
\end{figure}

In order to study the effects of each operator, we can turn on one Wilson coefficient coefficient $c_i$, corresponding to one operator $\mathbf{O}_i$, at a time, and set constraints on the $[M_\scalar, c_i]$ planes, where $M_\scalar$ is the scalar mass. We assume that only the third generation of SM fermions couples directly to the DM particle. The experimental data relevant for each operator is also shown in table~\ref{tab:operators-singlet}. The limits coming from different sources for each operator are summarized in figures.~\ref{fig:limits-1} and~\ref{fig:limits-2}.  We have computed these limits using the code \texttt{micrOMEGAs}~\cite{Belanger:2006is,Belanger:2013oya,Belanger:2018ccd}, and, for the direct detection limits on the $\mathbf{O}_G$ operator, eq.~(23) in ref.~\cite{Ruhdorfer:2019utl}. 

The top-left plot in figure.~\ref{fig:limits-1} corresponds to the renormalizable Higgs portal $\mathbf{O}_{\phi 1}$, studied in detail in ref.~\cite{Kanemura:2010sh,Djouadi:2011aa,Djouadi:2012zc}. The operators $\mathbf{O}_{\phi 4}$ and $\mathbf{O}_{\phi \square 1}$ also couple $\scalar$ only to the Higgs. The plots for the three Higgs operators show a downwards peak at $M = m_H/2$ in the curve for the abundance, due to the enhancement of annihilations at $s = 4 M^2$ through an s-channel Higgs. For the $\mathbf{O}_{\psi\phi 1}$ operators, the correct abundance is set by an approximately constant value of the Wilson coefficient, and only for DM masses above the mass of the corresponding fermion $\psi$. In general, for all the operators, the abundance line goes down quickly whenever a new annihilation channel opens up. The direct detection limits are relevant for the operators $\mathbf{O}_{\phi 1}$, $\mathbf{O}_{\phi 4}$, $(\mathbf{O}_{d \phi 1})_{33}$, $(\mathbf{O}_{u \phi 1})_{33}$ and $\mathbf{O}_G$. For all of them, these limits are stronger for masses between $10$ and $100$ GeV, where the Xenon1T experiment has the most sensitivity. The operators $\mathbf{O}_{\phi\square 1}$ and $\mathbf{O}_F$ contain derivatives and therefore give suppressed contributions to the DM-nucleon cross section at low energies,\footnote{$\mathcal{O}_{\phi \square 1}$ is the dominant operator in pseudo Nambu-Goldstone DM scenarios~\cite{Frigerio:2012uc, Ruhdorfer:2019utl}, which allows them to evade direct detection limits.} except for $\mathbf{O}_G$, which, despite the derivative suppression, gives a relatively large cross section with nucleons, because it couples the DM particles directly to gluons. $(\mathbf{O}_{e\phi 1})_{33}$ does not give any relevant contribution. The operators that couple $\scalar$ to the Higgs only are the ones that contribute to the invisible Higgs branching ratio. The operators coupling $\scalar$ only to gluons give monojet signals, through the process $gg,qq \to \scalar j$ \cite{Djouadi:2012zc}. In each plot, we include a red region were the EFT description is expected to break. This happens when the EFT expansion parameters $E/\Lambda$ are order 1,\footnote{Here, we are conservative, ignoring factors of $4\pi$ in loops contributions.} where $E$ represents the energy scale of the process under consideration, and $\Lambda = c^{-1/2}$ is the cut-off scale, for each Wilson 
coefficient $c$ of a dimension-6 operator. The relevant energy $E$ is the DM mass $M$ for all the observables we consider, except those from colliders, for which it is $E = v$. We thus require $M < \Lambda$ and $v < \Lambda$. The first condition induces the red regions, the second one translates into $c \lesssim \SI{15}{TeV^{-2}}$, which is always satisfied in our plots. 

From figures.~\ref{fig:limits-1} and~\ref{fig:limits-2}, one can see that direct detection constrains the mass of a DM particle coupling through the operators with coefficients $c_{\phi 1}$, $c_{\phi 4}$, $c_{d \phi 1}$ and $c_{u \phi 1}$ to be rather large, with the lower limit ranging from around $m_t \approx 175$ GeV for $c_{u\phi 1}$ to values close to $\SI{1}{TeV}$ for $c_{\phi 1}$. The growth with energy of the DM annihilation cross section for a derivative portal coupling with coefficient $c_{\phi \square 1}$ allows to have a DM mass just above the mass of the Higgs boson, $\gsim 125$ GeV. Below this mass value, the invisible Higgs branching ratio BR($H \to \text{inv})$ rules it out. The DM couplings to the gluon $c_G$ and $c_{\tilde{G}}$ are constrained to be smaller than about $\SI{3e-1}{TeV^2}$ by the CMS monojet searches for a DM mass below $\sim 200$ GeV. This, together with the constraint from the relic abundance, translates into a lower limit of $\sim \SI{20}{GeV}$ for the DM mass. Finally, the coefficients $c_{e\phi 1}$, $c_B$, $c_{\tilde{B}}$, $c_{W1}$ and $c_{\tilde{W}1}$ are practically unconstrained by experiment.

The leading-order dependence of each DM observable on the Wilson coefficients is easily obtained. All observables are functions of the squared amplitude $\left|\overline{\mathcal{M}}\right|^2$ for some process, which can be written as a quadratic function
\begin{equation}
  \left|\overline{\mathcal{M}}\right|^2 = \sum_{ij} f_{ij}(M_\scalar) c_i c_j,
  \label{eq:general-semianalytical}
\end{equation}
of the Wilson coefficients, with $M_\scalar$-dependent coefficients. Since $c_{\phi 1}$ is of order $1/\Lambda^0$ and the rest of the $c_i$ are of order $1/\Lambda^2$, each term in this sum can be of order $1/\Lambda^0$, $1/\Lambda^2$ or $1/\Lambda^4$. In what follows, we will keep all terms for completeness, but we remark that, when the dimension-4 operator is present, the precision of the formulas is $1/\Lambda^2$, since the interference between dimension-4 and dimension-8 operators would also be of order $1/\Lambda^4$. However, when the coefficient of the dimension-4 operator vanishes, the precision becomes $1/\Lambda^4$, since the lowest order to which dimension-8 operators can contribute is $1/\Lambda^6$.

Eq.~\eqref{eq:general-semianalytical} allows to obtain semi-analytical expressions for the main observables that we consider in our study. In order to do so, we need to find the $M_\scalar$ dependence of the $f_{ij}$ coefficients. We choose a simplified dependence that gives the correct results with less than $10\%$ error for most of the parameter space. For the annihilation cross section $\sigma_{\text{ann}}$ of two DM particles into two SM ones times the velocity in the non-relativistic limit, we choose:
\begin{equation}
  (\sigma_{\text{ann}}v)|_{s = 4M_\scalar^2}
  =
  \left(\SI{3e-26}{cm^3 s^{-1}}\right)
  \sum_{fij} \theta(2M_\scalar - m_f) P^f_{ij}(M_\scalar) A^f_{ij} \tilde{c}_i \tilde{c}_j,
  \label{eq:sigma-ann-semianalytical}
\end{equation}
with $\tilde{c}_{\phi 1} = c_{\phi 1}$, $\tilde{c}_i = c_i \cdot \si{TeV^2}$. The dimensionless coefficients $A^f_{ij}$ are $c_i$ and $M_\scalar$ independent. The $f$ index labels final states, while the $i$, $j$ indices label the Wilson coefficients. $m_f$ is the sum of the masses of the particles in the final state. Most of the mass dependence is encoded in the functions
\begin{equation}
  P_{ij}^f(M_\omega) = \left\{
    \begin{array}{ll}
      m_H^2 / M_\scalar^2
      & \text{for } f = H H \text{ and } i, j = \phi 1, \phi 4 \\
      1
      & \text{for } f = \text{bosons and } i = \phi 1, \phi 4; \; j = \phi \square 1, F, \widetilde{F} \\
      M_\scalar^2 / m_H^2
      & \text{for } f = \text{bosons and } i, j = \phi \square 1, F, \widetilde{F} \\
      \left(1 - \frac{m_\psi^2}{M_\scalar^2}\right)^{3/2}
      & \text{for } f = \text{fermions and }  i,j \neq \phi 1, \phi 4, \phi \square 1 \\
      \frac{m_H^4}{(4 M_\fermion^2 - m_H^2)^2 + \Gamma_{H \to b\bar{b}}^2 m_H^2}
      & \text{for } f = b\bar{b}, c\bar{c} \text{ and } i,j = \phi 1, \phi 4, \phi \square 1
    \end{array}
  \right.
\end{equation}
We obtain the coefficients $A_{ij}^f$ by evaluating $(\sigma_{\text{ann}} v)|_{s = 4M_\scalar^2}$ numerically using the code \texttt{micrOMEGAs}. The results are shown in table~\ref{tab:semianalytical}. As a general rule, the formula eq.~\eqref{eq:sigma-ann-semianalytical} can be trusted away from SM particle production thresholds $M_\scalar \simeq m_{\text{SM}}$ (with $m_{\text{SM}}$ the mass of any SM particle) and from the Higgs boson resonance $M_\scalar \simeq \frac12 m_H$.

The DM-nucleon scattering elastic cross section is simpler to parametrize. We find the approximate formula
\begin{equation}
  \sigma_N |_{s = (M_\fermion + m_N)^2}
  = (\SI{e-46}{cm^2}) \left(\frac{\SI{100}{GeV}}{m_N + M_\fermion}\right)^2
  \sum_{ij} N_{ij} \tilde{c}_i \tilde{c}_j.
\end{equation}
where the dimensionless coefficients $N_{ij}$ are $c_i$ and $M_\fermion$ independent. The sum runs over the $\phi 1$, $\phi 4$, $u \phi 1$ and $d \phi 1$ Wilson coefficients. Again, we obtain the values of $N_{ij}$ from the numerical calculation, and display them in table~\ref{tab:semianalytical}.

Similarly, the formula for the  branching ratio  of the Higgs boson decay into invisible DM particle is given by 
\begin{equation}
  \operatorname{BR}(H \to \text{inv})
  =
  \theta(m_H - 2 M_\fermion)
  \frac{
    \sqrt{1 - 4 M_\scalar^2 / m_H^2} \sum_{ij} H_{ij} c_i c_j
  }
  {
    1 + \sqrt{1 - 4 M_\scalar^2 / m_H^2} \sum_{ij} H_{ij} c_i c_j
  },
\end{equation}
where the $H_{ij}$ coefficients are $c_i$ and $M_\fermion$ independent. The sum
runs over the $\phi 1$, $\phi 4$ and $\phi \square$   Wilson coefficients. The values of the
$H_{ij}$ obtained from the numerical calculation are also given in
table~\ref{tab:semianalytical}.

\begin{table}
  \centering
  \begin{minipage}{0.49\textwidth}
    \centering
    $A^{H H}_{ij}$ \\[5pt]
    \begin{tabular}{c|ccc}
      \toprule
      $_i \backslash ^j$ & $\phi 1$ & $\phi 4$ & $\phi \square 1$ \\
      \hline
      $\phi 1$ & 390 & 150 & -56 \\
      $\phi 4$ &  & 16 &  -11 \\
      $\phi \square 1$ &  &  & 2.0 \\
    \end{tabular}
  \end{minipage}
  \begin{minipage}{0.49\textwidth}
    \centering
    $A^{\gamma \gamma}_{ij}$ \\[5pt]
    \begin{tabular}{c|ccccc}
      \toprule
      $_i \backslash ^j$ & $B$ & $\widetilde{B}$ & $W 1$ & $\widetilde{W} 1$ \\
      \hline
      $B$ & 9.1 & -12 & 350 & 0.43\\
      $\widetilde{B}$ & & 37 & -12 & 1400\\
      $W 1$ & & & 0.85 & 0.29\\
      $\widetilde{W} 1$ & & & & 3.4
    \end{tabular}
  \end{minipage}

  \vspace{10pt}
      
  \begin{minipage}{0.49\linewidth}
    \centering
    $A^{\gamma Z}_{ij}$ \\[5pt]
    \begin{tabular}{c|ccccc}
      \toprule
      $_i \backslash ^j$ & $B$ & $\widetilde{B}$ & $W 1$ & $\widetilde{W} 1$ \\
      \hline
      $B$ & 5.5 & -7.2 & 710 & 4.0\\
      $\widetilde{B}$ & & 22 & -8.1 & -2800 \\
      $W 1$ & & & 5.5 & 1.9 \\
      $\widetilde{W} 1$ & & & & 22
    \end{tabular}
  \end{minipage}
  \begin{minipage}{0.49\linewidth}
    \centering
    $A^{W^+ W^-}_{ij}$ \\[5pt]
    \begin{tabular}{c|ccccc}
      \toprule
      $_i \backslash ^j$ & $\phi 1$ & $\phi 4$ & $\phi \square 1$
      & $W 1$ & $\widetilde{W} 1$ \\
      \hline
      $\phi 1$ & 990 & 120 & -120 & 2.1 & 5.3  \\
      $\phi 4$ & & 3.6 & -7.4 & -0.057 & -0.057 \\
      $\phi \square 1$ & & & 3.9 & -6.9 & 6.2 \\
      $W 1$ & & & & 31 & 14 \\
      $\widetilde{W} 1$ & & & & & 120 \\
    \end{tabular}
  \end{minipage}

  \vspace{10pt}

  \begin{minipage}{0.8\linewidth}
    \centering
    $A^{ZZ}_{ij}$ \\[5pt]
    \begin{tabular}{c|ccccccc}
      \toprule
      $_i \backslash ^j$ & $\phi 1$ & $\phi 4$ & $\phi \square 1$
      & $B 1$ & $\widetilde{B} 1$ & $W 1$ & $\widetilde{W} 1$ \\
      \hline
      $\phi 1$ & 490 & 61 & -62 & 0.21 & 0.089 & 1.54 & -1.6 \\
      $\phi 4$ & & 1.8 & -3.4 & 0.15 & -0.028 & -0.028 & -0.028 \\
      $\phi \square 1$ & &  & 1.9 & -1.3 & -2.2 & -3.8 & 0.31 \\
      $B 1$ & & & & 0.83 & -1.1 & 360 & 6.1 \\
      $\widetilde{B} 1$ & & & & & 3.4 & 0.21 & 1400 \\
      $W 1$ & & & & & & 9.0 & 6.3 \\
      $\widetilde{W} 1$ & & & & & & & 36 \\
    \end{tabular}
  \end{minipage}

  \vspace{10pt}

  \begin{minipage}{0.25\textwidth}
    \centering
    $A^{GG}_{ij}$ \\[5pt]
    \begin{tabular}{c|ccc}
      \toprule
      $_i \backslash ^j$ & $G$ & $\widetilde{G}$ \\
      \hline
      $G$ & 120 & 500 \\
      $\widetilde{G}$ & & 33 \\
    \end{tabular}
  \end{minipage}
  \begin{minipage}{0.40\textwidth}
    \centering
    $A^{b \bar{b}}_{ij}$ \\[5pt]
    \begin{tabular}{c|ccc}
      \toprule
      $_i \backslash ^j$ & $\phi 1$ & $\phi 4$ & $\phi \square 1$ \\
      \hline
      $\phi 1$ & 34 & 4.1 & -1.7 \\
      $\phi 4$ &  & 0.12 & -0.10 \\
      $\phi \square 1$ &  &  & 0.2 \\
    \end{tabular}
  \end{minipage}
  \begin{minipage}{0.25\textwidth}
    \centering
    $A^{c \bar{c}}_{ij}$ \\[5pt]
    \begin{tabular}{c|ccc}
      \toprule
      $_i \backslash ^j$ & $\phi 1$ & $\phi 4$ \\
      \hline
      $\phi 1$ & 2.5 & 0.30 \\
      $\phi 4$ & & 0.0090
    \end{tabular}
  \end{minipage}

  \vspace{10pt}

  $A^{t\bar{t}}_{u\phi 1,u\phi 1} = 11$, \qquad
  $A^{b\bar{b}}_{d\phi 1,d\phi 1} = 11$, \qquad
  $A^{\tau^+\tau^-}_{e\phi 1,e\phi 1} = 3.8$.

  \vspace{10pt}
  
  \begin{minipage}{0.48\textwidth}
    \centering
    $N_{ij}$ \\[5pt]
    \begin{tabular}{c|cccc}
      \toprule
      $_i \backslash ^j$ & $\phi 1$ & $\phi 4$ & $d\phi 1$ & $u\phi 1$ \\
      \hline
      $\phi 1$ & 36000 & 4300 & -10000 & -260 \\
      $\phi 4$ & -- & 130 & -600 & -16 \\
      $d\phi 1$ & -- & -- & 700 & 37 \\
      $u\phi 1$ & -- & -- & -- & 0.49
    \end{tabular}
  \end{minipage}
  \begin{minipage}{0.49\textwidth}
    \centering
    $H_{ij}$ \\[5pt]
    \begin{tabular}{c|ccc}
      \toprule
      $_i \backslash ^j$ & $\phi 1$ & $\phi 4$ & $\phi\square 1$ \\
      \hline
      $\phi 1$ & 3189 & 387.1 & -97.20 \\
      $\phi 4$ & -- &  11.72 & -6.043 \\
      $\phi \square 1$ & -- & -- & 0.7788
    \end{tabular}
  \end{minipage}
  \caption{Coefficients $A^{fg}_{ij}$, $N_{ij}$ and $H_{ij}$ in the semi-analytical formulas for the DM annihilation cross section, DM-nucleon scattering cross section and Higgs to invisible branching ratio.}
  \label{tab:semianalytical}
\end{table}

\begin{figure}[!h]
  \centering
  \includegraphics[width=0.98\textwidth]{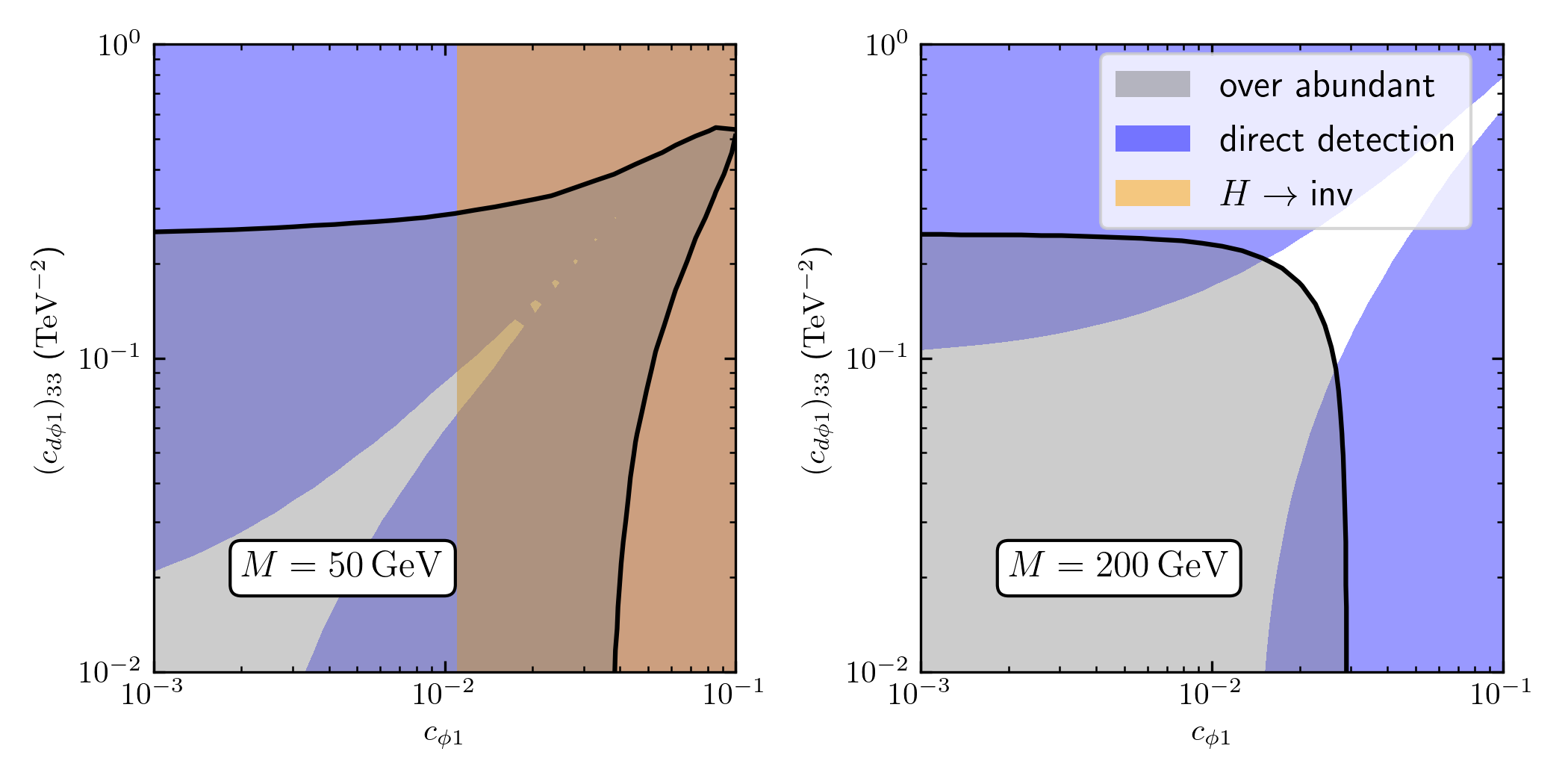}
  \caption{Line of points (in black) that give the correct DM abundance, together with the region that gives an overabundance of DM (in gray), the region excluded by direct-detection experiments (in blue) and the region excluded by the upper bound on $\operatorname{BR}(H \to \text{inv})$ (in orange), in the space of $c_{d \phi 1}$ vs $c_{\phi 1}$, for different values of $M_\scalar$. 
  }\label{fig:cphi1-vs-cdphi1}
\end{figure}

These expressions greatly facilitate the exploration of the parameter space of the Wilson coefficient. For example, examining the table of $N_{ij}$ coefficients suggests that a cancellation between the contributions of the operators with $c_{\phi 1}$ and $c_{d \phi 1}$ coefficients to the DM-nucleon cross section can occur. This opens the possibility of evading the stringent limits from direct detection.

We show in figure.~\ref{fig:cphi1-vs-cdphi1} the limits in the $[c_{\phi 1}, c_{d \phi 1}]$ plane for two values of the DM particle mass, $M_\scalar=50$ and 200 GeV, computed numerically using \texttt{micrOMEGAs}. As can be seen there, the semi-analytical treatment has led us to a viable possibility: while masses below the TeV scale are excluded by direct detection limits when only the dimension-4 portal is present, the addition of a dimension-6 operator allows to avoid them while keeping the correct relic DM abundance. For $M_\scalar > \frac12 m_H$, no other limits apply. For $M_\scalar < \frac12 m_H$, the tiny set of values not excluded by direct detection is only ruled out by the limits on the invisible Higgs branching ratio $\operatorname{BR}(H \to \text{inv})$. Similar results as those shown in figure.~\ref{fig:cphi1-vs-cdphi1}, with in particular a suppression of the direct detection limits,  have been obtained in a different context in ref.~\cite{Alanne:2020xcb}. 


\section{Lepton doublet dark matter}
\label{sec:lepton-example}

In this section, we consider the possibility in which the DM particle belongs to a non-singlet $\SU(2)_L \times \U(1)_Y$  multiplet  and illustrate the use of our approach in the specific case of a DM lepton doublet.\footnote{There is a vast literature on DM scenarios with singlet-doublet leptons and, in the context of Higgs portal models, they have been reviewed in e.g. ref.~\cite{Arcadi:2019lka}. Doublet leptons alone have been discussed in the context of supersymmetric models with higgsino DM, see for instance refs.~\cite{Drees:1996pk,Djouadi:2005gj,Baer:2013vpa}, and in the non-supersymmetric case, for instance in refs.~\cite{Joglekar:2012vc,Dedes:2016odh}. Non-singlet scalar multiplets have been explored in the context of composite Higgs models in~\cite{Carmona:2015haa,Ballesteros:2017xeg}.}  We start by discussing the contributions of the various operators to the DM annihilation cross sections and to the mass mass-splitting. 

Non-singlet multiplets are fundamentally different from singlets in two ways: they have dimension-4 couplings fixed by gauge invariance and more than one particle is present in the spectrum near the DM mass. In order to illustrate this possibility, we consider the case in which the multiplet is a spin-1/2 doublet $\Delta = (N, E)^T$ with hypercharge $Y = -1/2$, so that $\widetilde{\Delta}$ corresponds to the $Y = 1/2 > 0$ multiplet $\chi$ we considered in section~\ref{sec:eft}. We choose the opposite hypercharge in order for $\Delta$ to have the same quantum numbers as the usual SM lepton doublet. The leading higher-dimensional operators for $\Delta$ have dimension 5 and are listed in table~\ref{tab:lepton-doublet-operators}.

\begin{table}[!h]
\renewcommand{\arraystretch}{1.25}
    \centering
    \begin{tabular}{cc}
        \toprule
        Operators
        & Constraints \\
        \midrule
        $\mathbf{O}_{\phi 1}$, $\mathbf{O}_{\phi 2}$,
        $\mathbf{O}_{\phi 3}$, $\mathbf{O}_{\phi 4}$
        &  $H \to \text{inv}$, direct detection \\
        $\mathbf{O}_B$, $\mathbf{O}_W$
        & -- \\
        \bottomrule
    \end{tabular}
    \caption{Dimension-5 operators for a DM lepton doublet. All of them contribute to the DM annihilation cross section.}
    \label{tab:lepton-doublet-operators}
\end{table}

Apart from the vector-like mass term in eq.~(\ref{eq:quad-lag-fermion}), the particles contained in the $\Delta$ multiplet receive tree-level corrections to their mass from the four dimension-5 $\mathbf{O}_{\phi i}$ operators. After the spontaneous breaking of the electroweak symmetry, the mass Lagrangian takes the form
\begin{align}
    -\mathcal{L}_{\text{mass}}
    &=
    \begin{pmatrix}
        \overline{N^c_L} & \overline{N}_R
    \end{pmatrix}
    \begin{pmatrix}
        v^2 c_{\phi 3} 
        & \frac{1}{2}\left(M - \frac{v^2}{2} (c_{\phi 1} - c_{\phi 2})\right)
        \\
        \frac{1}{2}\left(M - \frac{v^2}{2} (c_{\phi 1} - c_{\phi 2})\right)
        &
        v^2 c_{\phi 4}
    \end{pmatrix}
    \begin{pmatrix}
        N_L \\ N_R^c
    \end{pmatrix}
    \nonumber
    \\
    &\phantom{=}
    +
    \left(
        M - \frac{v^2}{2} (c_{\phi 1} + c_{\phi 2})
    \right) \overline{E}_L E_R    
    + \text{h.c.} \, . 
\end{align}
The diagonalization of the mass matrix for the states $N_L$ and $N_R^c$ is performed by means of the unitary transformation
\begin{equation}
    \begin{pmatrix}
        N_L \\ N_R^c
    \end{pmatrix}
    =
    \begin{pmatrix}
        \cos\theta & -e^{i\phi} \sin\theta \\
        e^{-i\phi} \sin\theta & \cos\theta
    \end{pmatrix}
    \begin{pmatrix}
        e^{-i\alpha} & 0 \\
        0 & e^{-i\beta}
    \end{pmatrix}
    \begin{pmatrix}
        N_1 \\ N_2
    \end{pmatrix},
    \label{eq:diagonalization}
\end{equation}
with mixing angle $\theta$ given by
\begin{equation}
    \sin^2 \theta = 
    \frac{1}{2}
    + \frac{|c_{\phi 3}|^2 - |c_{\phi 4}|^2}{2}
    \left\{
        \left(|c_{\phi 3}|^2 - |c_{\phi 4}|^2\right)^2
        + |c_{\phi 3} + c_{\phi 4}|
        \left|
            \frac{M}{2 v^2} - \frac{c_{\phi 1} - c_{\phi 2}}{4}
        \right|
    \right\}^{-1} \, .
\end{equation}
After diagonalization, we find that our multiplet contains three particles: two heavy Majorana neutrinos $N_1$ and $N_2$, and a heavy charged lepton $E$ with masses given by
\begin{align}
    M_{N_1} &= M \left[
        1
        + \frac{v^2}{2 M} \left(
            -\re c_{\phi 1} + \re c_{\phi 2} - 2 |c_{\phi 3} + c_{\phi 4}^*|
        \right)
        + O\left(c_i^2\right)
    \right], \\
    M_{N_2} &= M \left[
        1
        + \frac{v^2}{2 M} \left(
            -\re c_{\phi 1} + \re c_{\phi 2} + 2 |c_{\phi 3} + c_{\phi 4}^*|
        \right)
        + O\left(c_i^2\right)
    \right],
    \\
    M_E &= M \left[
        1 
        - \frac{v^2}{2 M} \left(
            \re c_{\phi 1} + \re c_{\phi_2}
        \right)
        + O\left(c_i^2\right)
    \right] \, . 
\end{align}
In order for the neutral state $N_1$ to be stable, its mass must be smaller than the charged lepton $E$ mass: $M_{N_1} < M_E$. This occurs if and only if
\begin{equation}
  \re c_{\phi 2} < |c_{\phi 3} + c_{\phi 4}^*|.
  \label{eq:splitting-condition}
\end{equation}
When both $c_{\phi 3}$ and $c_{\phi 4}$ vanish, the masses $M_1$ and $M_2$ become degenerate, and it is then convenient to keep the neutrino mass matrix anti-diagonal, and view $N_L$ and $N_R$ and as the left- and right-handed components of a unique Dirac neutrino.

The coefficients $c_{\phi 2}$, $c_{\phi 3}$ and $c_{\phi 4}$ thus provide a general order-$1/\Lambda^2$ parametrization of new physics effects in the mass splitting. Even when no new physics is present, there will be a remaining splitting induced by loops involving the gauge interactions~\cite{Cirelli:2005uq,Cirelli:2009uv,Bottaro:2021srh}. We will assume that the tree-level effects of the dimension-5 coefficients are dominant here, and neglect the loop-induced splitting.

Let us briefly summarize the present experimental constraints on the masses and couplings of the vector-like leptons. First, because of the $\mathbb{Z}_2$ symmetry which has been introduced to stabilize the DM particle, these heavy leptons will not mix with the SM ones and, thus, will not decay into SM particles and cannot be produced in association with them. Hence, the usual stringent constraints on heavy leptons from the anomalous magnetic moments of the muon or electron~\cite{Criado:2021qpd} or from the associated production with charged SM leptons and neutrinos at LEP2 \cite{Tanabashi:2018oca} will not hold\footnote{Some of the dimension-6 operators in table~\ref{tab:fermion-basis} would contribute to the lepton anomalous magnetic moments; however, they are not relevant for our discussion here.}. In the case of the DM particles, the only relevant constraint will come from the invisible decays of the $Z$ boson into DM pairs (when the particle has non-zero hypercharge) which sets a bound of $M_{N_1} \gsim 45$ GeV in the  Dirac and $M_{N_1} \gsim 39.5$ GeV in the Majorana cases \cite{Tanabashi:2018oca}; there are also bounds from the invisible decay of the Higgs boson, $M_{N_1} \lsim \frac12 m_H= 62.5$ GeV \cite{Aaboud:2019rtt,Sirunyan:2018owy} if the Higgs-$N_1N_1$ coupling is not too small.  In the case of the charged lepton,  there is a strict lower bound on its mass from the pair production at LEP2: $M_E \gsim 103$ GeV for an almost stable $E$ and $M_E \gsim 101$ GeV if it decays into a light neutrino and a $W$ boson \cite{Tanabashi:2018oca}. This bounds translates to a bound on $M_{N_1}$ when the leptons are close in mass. 

Searches of these particles have also been conducted at the LHC where they can be produced in the Drell-Yan type processes $p \to E^+ E^-, NN, E^\pm N$ ($N$ stands for a generic neutral lepton)  with the latter mode being by far dominant as it is mediated by the exchange of the charged $W$ boson \cite{delAguila:2007qnc}.  However, the constraints are again tight only in the case where the heavy leptons mix with the ordinary ones, leading to prompt electrons or muons and $W,Z$ bosons in the final state \cite{Aad:2020fzq,Sirunyan:2019bgz}. In our case, the lightest $N$ particle is stable while the next-to-lightest $N'$ and the charged lepton $E$ states will decay into the DM and $Z$ or $W$ bosons. If the mass difference between the DM particle and its companions is very small, as it is necessary the case for the co-annihilation mechanism to be relevant, the intermediate gauge bosons would be far off-shell and the available phase space would be tiny, so the processes would correspond to  the  production  of  long-lived  particles that lead  to displaced vertices or particles being eventually stable at the detector level. There are searches by ATLAS and CMS of stable and long-lived charged sleptons and charginos, which constrain these particles to have masses above a few hundred  GeV, see e.g. refs.~\cite{Aad:2015qfa,Aaboud:2019trc}, and could be relevant in our case. However, as these bounds are model dependent and are very sensitive to the details of the analyses, we will not include them in our study. Dedicated experiments such as MoEDAL, FASER or  SHiP for instance could probe long-lived particles more efficiently, see e.g. ref.~\cite{Mitsou:2020okk} for a review.

In our analysis, we will therefore simply use the constraints $M_E \gsim 100$ GeV for the charged leptons and $M_N \gsim 45$ GeV for the neutral ones, which have a non-zero hypercharge. We will of course also take into account the bounds from the Higgs invisible branching ratios as measured at the LHC, BR($H \to NN) \lsim 20\%$ \cite{Aaboud:2019rtt,Sirunyan:2018owy} as well as constraints from direct and  indirect detection of the DM particle. 

According to eq.~(\ref{eq:splitting-condition}), there are two classes of Wilson coefficients, those that induce a splitting between the masses $M_E$ and $M_{N_1}$, and those that do not, which we denote respectively by $c_M$ and $c_{\slashed{M}}$:
\begin{equation}
    c_M \in \{c_{\phi 2}, c_{\phi 3}, c_{\phi 4}\},
    \qquad
    c_{\slashed{M}} \in \{c_{\phi 1}, c_B, c_W\}.
\end{equation}
An analysis similar to the one in section~\ref{sec:scalar}, of the effects of each coefficient individually while turning off all others, is only possible for the $c_M$ coefficients. 

The relative effects of gauge interactions and dimension-5 operators in the DM annihilation cross section are controlled by two parameters: the relative mass splitting of the charged and lightest neutral leptons $\Delta \equiv (M_E - M_{N_1}) / M_{N_1} \sim c_M v^2 / M_{N_1}$, and the effective Yukawa coupling of the neutral lepton $Y_{N_1} \sim c_M v$. Then, there are three different cases to consider:
\begin{enumerate}
\item[$i)$] $\Delta \lesssim 0.1$: in which the co-annihilation processes dominate in the cross sections. In terms of $c_M$ and $M_{N_1}$ this happens when
  \begin{equation}
    c_M \lesssim (\SI{0.02}{TeV^{-1}}) \left(\frac{M}{\SI{10}{GeV}}\right)
  \end{equation}
 
\item[$ii)$]  $\Delta \gtrsim 0.1$ and $Y_{N_1} \lesssim g_{\text{EW}}$: in which DM annihilation processes through the $s$-channel exchange of the $Z$-boson dominate. This occurs is when
  \begin{equation}
    (\SI{0.02}{TeV^{-1}}) \left(\frac{M}{\SI{10}{GeV}}\right)
    \lesssim c_M \lesssim \SI{3}{TeV^{-1}}.
  \end{equation}
    
\item[$iii)$] $Y_{N_1} \gtrsim g_{\text{EW}}$: the annihilation processes through the $s$-channel exchange of the Higgs boson (with dim-5 operators) dominate, when
  \begin{equation}
    \SI{3}{TeV^{-1}} \lesssim c_M \, .
  \end{equation}

\end{enumerate}
There are two issues with case $iii)$: first, the masses $M_{N_1}$ that correspond to the correct abundance are low, about $\SI{10}{GeV}$ for $M$, and thus excluded by LEP searches; and second, the perturbative expansion of the EFT is broken if dim-5 interactions have a similar size to dim-4 ones. $c_M$ being high also means that the masses $M_{N_1}$ and $M_E$ should be small, $M_{N_1}, M_E \! \lesssim \! \Lambda \! \sim \! c_M^{-1}$. We conclude that in almost all of the allowed parameter space, dimension-5 operators affect the abundance only through the splitting.

Not all observables depend on the $c_M$ coefficients through the splitting. The case of direct detection, for instance, can be strikingly different since, as we have discussed in detail in section~\ref{sec:Z-couplings}, the couplings to the $Z$ boson leading to spin-independent direct detection can be naturally suppressed in this case~\cite{Chala:2015ama}. Indeed,
when $N_1$ is a Majorana fermion, its vector current vanishes:
\begin{align}
    \overline{N}_1 \gamma^\mu N_1 
    = \overline{N^c_1} \gamma^\mu N^c_1
    = N^T_1 C^\dagger \gamma^\mu C \overline{N}^T_1
    = - \overline{N}_1 \gamma^\mu N_1
    \label{eq:vector-current}
\end{align}
This means that its coupling to the $Z$ boson must be through the axial vector current:
\begin{align}
    \mathcal{L}_{NZ}
    &= \frac{e}{c_W s_W} \bar{N} \gamma^\mu N Z_\mu
    \nonumber
    \\
    &\supset 
    \frac{e}{c_W s_W}
    \bar{N}_1 
    (e^{i\alpha} \cos\theta P_R - e^{-i(\phi + \alpha)} \sin\theta P_L) 
    \gamma^\mu 
    (e^{-i \alpha} \cos\theta P_L - e^{i(\phi + \alpha)} \sin\theta P_R) 
    N_1 
    Z_\mu
    \nonumber
    \\
    &= 
    - i g_Z \bar{N}_1 \gamma^\mu \gamma^5 N_1 Z_\mu,
\end{align}
where we have used eq.~\eqref{eq:diagonalization} in passing from the first to the second line, eq.~\eqref{eq:vector-current} in going from the second to the third one,
$s_W$ and $c_W$ are the sine and cosine of the Weinberg angle, and
\begin{equation}
    g_Z = \frac{e \cos(2\theta)}{2 s_W c_W}.
\end{equation}
Thus, this interaction vanishes for the value $\theta = \pi / 4$ of the mixing angle,  which occurs if $c_{\phi 3} = c_{\phi 4}$. In the case in which $c_{\phi 3} > 0$, $c_{\phi 4} = 0$ and $M/c_{\phi i} \gg v^2$, the mixing angle and the $N_1N_1Z$ coupling are simply given by
\begin{equation}
    \sin^2 \theta \simeq \frac{1}{2} + \frac{v^2 |c_{\phi 3}|}{M} \, , \ \ \  
    g_Z \simeq \frac{e v^2 |c_{\phi 3}|}{c_W s_W M}
    \simeq 0.04 \left(\frac{c_{\phi 3}}{\SI{0.1}{TeV^{-1}}}\right)
    \left(\frac{\SI{100}{GeV}}{M}\right).
\end{equation}
The coupling of the $N_1$ states to the Higgs boson would be given, in this case, by
\begin{equation}
    \mathcal{L}_{N_1 N_1 H} = Y_{N_1} \overline{N_1} N_1 H,
    \qquad Y_{N_1} = \sqrt{2} v c_{\phi 3} 
    \simeq 0.03 \left(\frac{c_{\phi 3}}{\SI{0.1}{TeV^{-1}}}\right) \, .
\end{equation}
Since the $N_1$ coupling to the $Z$ boson only contributes to the spin-dependent DM-nucleon cross section while the coupling to the $H$ boson contributes to the spin-independent one, and the limits on the spin-dependent cross section are at least five orders of magnitude weaker~\cite{Aprile:2019dbj}, the direct detection limits are essentially controlled by the interaction with the Higgs particle. 

\begin{figure}[!h]
\vspace*{-2mm}
    \centering
    \includegraphics[width=0.98\textwidth]{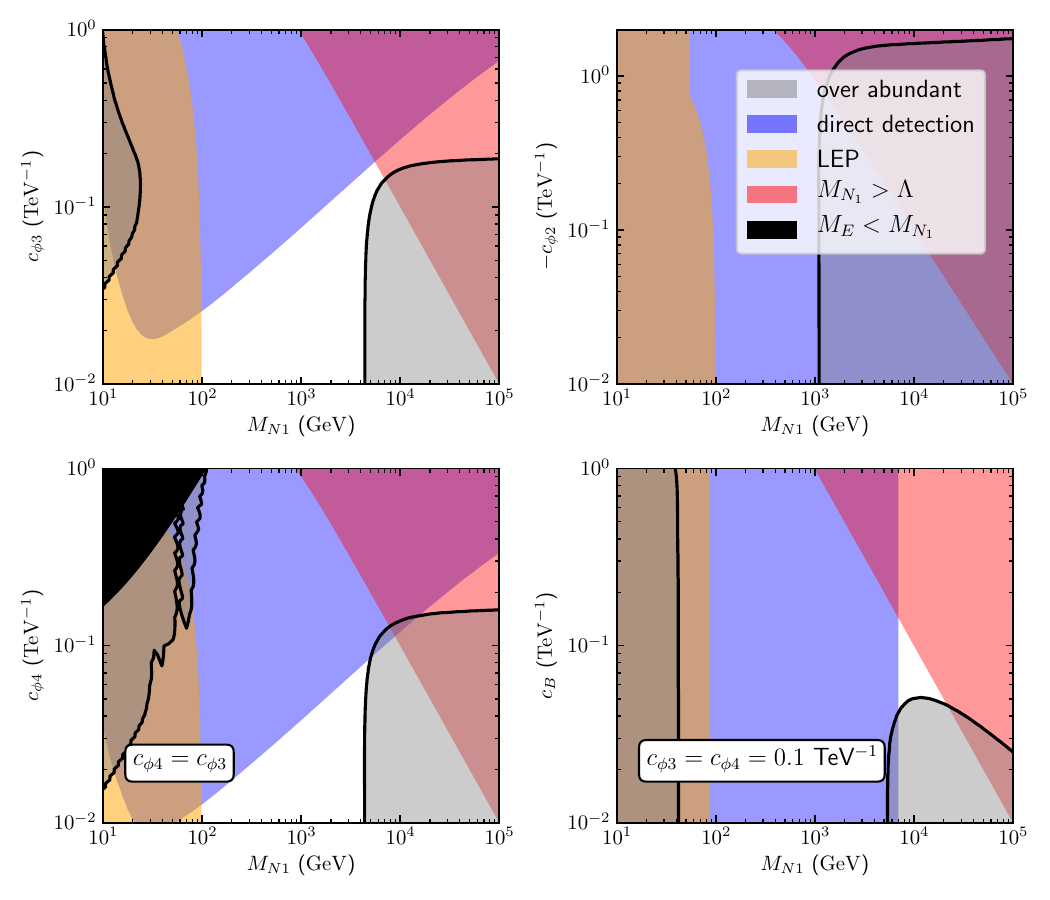}
\vspace*{-2mm}
    \caption{Constraints on the coefficients of operators for the DM lepton doublet. The points over the solid black line give the correct DM abundance, while the gray region gives an overabundance of DM. The orange region is constrained by LEP searches of heavy leptons. The region excluded by direct-detection experiments is shown in blue. Indirect detection places not relevant limits.}
    \label{fig:doublet-panel}
\vspace*{-2mm}
\end{figure}

In figure.~\ref{fig:doublet-panel}, we show the limits in various regions of parameter space from the relevant experimental constraints: the DM abundance, the direct detection cross section and the searches for heavy leptons at LEP. We also include in each case a red area in which the EFT description is expected to break down, which is the region where $M_E > \Lambda = c^{-1}$ for each Wilson coefficient $c$. The two plots at the top correspond to turning on one $c_M$-type coefficient at a time: $c_{\phi 3}$ and $c_{\phi 2}$. We choose negative values for $c_{\phi 2}$, as required by eq.~\eqref{eq:splitting-condition} when $c_{\phi 3} = c_{\phi 4} = 0$. On the bottom plots we turn on more than one coefficient at a time. On the left, we display the limits in $[M_{N_1}, c_{\phi 3}]$ plane when the relation $c_{\phi 4} = c_{\phi 3}$ is imposed. On the right, we consider the case in which $c_{\phi 4} = c_{\phi 3} = \SI{0.1}{TeV^{-1}}$, and explore the $[M_{N_1}, c_B]$ space.

The generation of the correct abundance is dominated by co-annihilations for the black curve on the right of each plot in figure.~\ref{fig:doublet-panel}. This corresponds to the regime labeled $i)$ above. The curve on the left is dominated by annihilations through the $Z$, corresponding to case $ii)$. The limit from LEP becomes just the upper bound on $M_{N_1}$ for large $c_M$, since this implies a large splitting; and just the upper bound on $M_E$, for small $c_M$. In the three plots in which $c_{\phi 3} \neq 0$, $N_1$ is a Majorana particle, and the $N_1N_1Z$ coupling does not contribute to the spin-independent DM-nucleon cross section. Thus, the region excluded by direct detection is controlled by the $N_1N_1H$ coupling, which is proportional to the relevant $c_M$ coefficient. On the top-right plot, the $N_1$, $N_2$ pair is degenerate and can be seen as a single Dirac fermion. Then, there is an spin-independent contribution to the DM-nucleon cross section from the $N_1N_1Z$ coupling, leading to strong limits from direct detection.

Since the $c_{\phi 3}, c_{\phi 4} \to 0$ limit should lead to the Dirac case, one may wonder how the strong limits from direct detection arise in that case. In this limit, the mass splitting between $N_1$ and $N_2$ becomes small. Then, the collision between $N_1$ and a nucleon may produce an $N_2$ particle through the exchange of a $Z$ boson with vector (not axial vector) couplings, thus contributing to the spin-independent cross section. However, in order for this to happen, the mass splitting must be smaller than the typical kinetic energy of a DM particle near the Earth, which is $M v^2 / 2$, with $v \simeq \num{1e-3}$. This does not happen in the region we consider.

On the bottom-right plot of figure.~\ref{fig:doublet-panel}, we see that most of the limits on $[M_{N_1}, c_B]$ space do not depend on $c_B$. This is because its effects are subdominant in each case: the annihilation cross section on the left is dominated by the dimension-4 coupling to the $Z$, the mass is controlled by $M$ and $c_M$, and the contribution of $c_B$ to direct detection is suppressed by the DM velocity while that of the $c_M$ is not. However, for high-enough masses, $c_B$ has a relevant contribution to the annihilation cross section, since then the effective coupling $M c_B$ becomes comparable to the dimension-4 gauge couplings and larger than the effective $v c_M$ coupling from the other coefficients.


\section{Conclusions}
\label{sec:conclusions}

Effective field theory has proven to be a useful approach for model independent studies of the  phenomenology of the weakly interacting, massive and cosmologically stable particles that are expected to form the dark matter in the universe. In this paper, we have investigated the possibility of going beyond the simple EFTs which have been discussed in the past, in which the interaction of the DM particles with the SM ones is parameterized in terms of a single or a few dominant operators, renormalisable or not, with the mediators of the interactions assumed to be very heavy and integrated out.  Assuming that the DM particles have spin-0, $\frac12$ or 1, and appear as the neutral components of a single $\SU(2)_L \times \U(1)_Y$ multiplet with arbitrary isospin and hypercharge, we construct a general and non-redundant basis for all relevant gauge-invariant operators in this theory up to dimension six. This complete EFT can be viewed as being simply the SMEFT that has been widely discussed in recent years, in which an additional scalar, vector or fermionic multiplet containing the DM particle is added to the SM spectrum.

We have then illustrated the usefulness of such a general approach with two specific examples. We have first considered a singlet scalar DM state and constrained the Wilson coefficients of all the relevant operators using present data on the DM: its cosmological relic abundance, direct and indirect detection in astroparticle experiments and searches in collider experiments, like through the invisible decays of the SM Higgs boson at the LHC. In order to simplify such an analysis, we have proposed a rather simple and convenient set of semi-analytical expressions for these DM observables, which allows to explore the complex parameter space  of the various operators in a very efficient manner. Using our expressions it is, for instance, trivial to find possible cancellations in the contribution to very constrained observables, like direct DM detection, thus opening up the allowed parameter space.

In a second illustration, we have considered the example of a vector-like lepton isodoublet, and we have discussed the interplay between the gauge interactions that are present in this case and dimension-5 operators. In particular, we have studied the simultaneous impact of some the higher dimension operators which enter both the mass splitting between the different members of the multiplet and the annihilation and co-annihilation cross sections of these states into SM particles. We have also described how, when the DM particle is Majorana, its gauge coupling to the $Z$ boson leads to a vanishing contribution to the very constrained spin-independent DM-nucleon cross-section relevant for direct detection experiments. In this case the direct detection constraints are dominated by higher-dimensional operators despite the fact that the DM particle has non-zero hypercharge. We have shown that this is a quite generic mechanism, which can be applied to scalar, fermion and vector DM particles with non-vanishing hypercharge, equal to $Y=1/2$ if we restrict ourselves to operators of mass dimension up to six, as we have done in this work.

\vspace{5mm}
\noindent \textbf{Acknowledgement.} 

\noindent Discussions with Giorgio Arcadi, Mikael Chala, Guilherme Guedes, Ennio Salvioni, Javi Virto and Jure Zupan are gratefully acknowledged.  J.C.C. is supported by the STFC under grant ST/P001246/1. The work of A.D. is supported by the Junta de Andalucia through the Talentia Senior program and, in part, by the ERC Mobilitas Plus grant MOBTT86. This work has been partially supported by the Ministerio de Ciencia e Innovaci\'on project PID2019-106087GB-C22 and by Junta de Andalucia projects FQM-101, A-FQM-211-UGR18, P18-FR-4314 and SOMM17/6104/UGR (including ERDF).

\clearpage

\appendix

\section{Operator basis}
\label{app:bases}

In this appendix, we present the basis of operators of dimension $D \leq 6$ containing two multiplets $\mathcal{X}$, for the EFT defined in section~\ref{sec:eft}. The operators for scalar multiplet $\scalar$ are shown in tables~\ref{tab:scalar-basis-bosons} and~\ref{tab:scalar-basis-fermions}; for a fermion $\fermion$, they are in table~\ref{tab:fermion-basis}; and for a vector $\vector$, they are in tables~\ref{tab:vector-basis-tensors}, \ref{tab:vector-basis-fermions} and~\ref{tab:vector-basis-higgs}. We do not list the complex conjugate of non-hermitian operators. We have checked the counting of operators with each field content for the lowest isopins $T = 0, 1/2, \ldots, 4$ using the code \texttt{BasisGen}~\cite{Criado:2019ugp}. In all the tables use the following notation:
\begin{itemize}
    \item The $\SU(2)$ index of $\mathcal{X}=\scalar,\fermion,\vector$, any $\SU(2)$ doublet index  and Lorentz spinor indices are implicit.
    \item Lower case letters $a$, $b$, \ldots are used for $\SU(2)$ triplet indices, capital letters $A$, $B$, \ldots are used for $\SU(3)$ octet indices and the letters $I$, $J$, \ldots are used for $\SU(2)$ quadruplet indices.
    \item $T^a$ denotes the $\SU(2)$ generators in the $\mathcal{X}$ representation, $\epsilon$ is the Levi-Civita symbol, $C_{Iab}$ denotes the quadruplet-triplet-triplet $\SU(2)$ Clebsh-Gordan coefficients. When the $\SU(2)$ isospin $T$ is larger than $1/2$, we denote by $Q^a$ the unique set of square matrices acting on $\mathcal{X}$ such that $\mathcal{X}^\dagger Q^a \mathcal{X}$ transforms as a quadruplet.
    \item The tilde symbol\  $\widetilde{\phantom{\cdot}}$\  is used to denote both the operation defined in eq.~(\ref{eq:SU2-conjugation}), when used over the DM multiplet and the dual $\widetilde{F}_{\mu\nu}=\frac{1}{2}\epsilon_{\mu\nu\rho\sigma}F^{\rho\sigma}$ when applied to a field-strength tensor $F$.  \item $G^A_{\mu\nu}$, $W^a_{\mu\nu}$ and $B_{\mu\nu}$ are, respectively, the $\SU(3)_C$, $\SU(2)_L$ and $\U(1)_Y$ field strengths. The SM matter content is denoted by $\phi$ for the Higgs doublet, $q$ and $l$ for the left-handed quark and lepton doublets and $u$, $d$ and $e$ for the charge 2/3 and -1/3 quarks and charged lepton singlets.
    \item We define the standard linear combinations of covariant derivatives $\DLR{\mu}{} = D_\mu - \overset{\leftarrow}{D}_\mu$ for both any multiplet, $\DLR{\mu}{a} = T^a D_\mu  - \overset{\leftarrow}{D}_\mu T^a$ when applied to $\mathcal{X}$ and $\DLR{\mu}{a} = \sigma^a D_\mu - \overset{\leftarrow}{D}_\mu \sigma^a $ when applied to $\phi$.
\end{itemize}

The allowed operators for real representations are those for which both the ``$\SU(2)$ irrep'' and the ``Hypercharge'' columns of tables~\ref{tab:scalar-basis-bosons}--\ref{tab:vector-basis-fermions} read ``any''. In these tables, the operators are listed assuming a complex multiplet. The real case is recovered by identifying $\mathcal{X} = \mathcal{X}^\dagger$. For spinors this means $\fermion_L = \fermion_R^c$, so that some operators are duplicated and the copies are to be discarded. Moreover, for vector fields the operators containing $\rho^\dagger_\mu \rho_\nu$ contracted with an anti-symmetric tensor $\sigma^{\mu\nu}$, $F^{\mu\nu}$ or $\widetilde{F}^{\mu\nu}$ will vanish.

\clearpage

\subsection{Operators for a scalar multiplet $\scalar$}

\begin{table}[!h]
\renewcommand{\arraystretch}{1.25}
  \centering
  \begin{tabular}{cccccc}
    \toprule
    Name
    & Operator
    & SU(2) irrep
    & Hypercharge
    & Dimension
    \\
    \midrule
    $\mathbf{O}_{\phi 1}$
    & $(\scalar^\dagger \scalar) (\phi^\dagger \phi)$
    & any
    & any
    & 4
    \\
    $\mathbf{O}_{\phi 2}$
    & $(\scalar^\dagger T^a \scalar) (\phi^\dagger \sigma^a \phi)$
    & $T > 0$
    & any
    & 4
    \\
    $\mathbf{O}_{\phi 3}$
    & $(\scalarTilde^\dagger T^a \scalar) (\phi^\dagger \sigma^a \widetilde{\phi})$
    & $T\in \mathbb{Z} + 1/2$
    & $1/2$
    & 4
    \\
    $\mathbf{O}_{\phi 4}$
    & $(\scalar^\dagger \scalar) (\phi^\dagger \phi)^2$
    & any
    & any
    & 6
    \\
    $\mathbf{O}_{\phi 5}$
    & $(\scalar^\dagger T^a \scalar) (\phi^\dagger \phi) (\phi^\dagger \sigma^a \phi)$
    & $T > 0$
    & any
    & 6
    \\
    $\mathbf{O}_{\phi 6}$
    & $i\epsilon_{abc}
      (\scalar^\dagger T^a \scalar)
      (\phi^\dagger \sigma^b \phi)
      (\phi^\dagger \sigma^c \phi)$
    & $T > 1/2$
    & any
    & 6
    \\
    $\mathbf{O}_{\phi 7}$
    & $(\scalarTilde^\dagger T^a \scalar) (\phi^\dagger \sigma^a \widetilde{\phi})
      (\phi^\dagger \phi)$
    & $T > 0$
    & $1/2$
    & 6
    \\
    \midrule
    $\mathbf{O}_{\phi \square 1}$
    & $(\scalar^\dagger \scalar) \square (\phi^\dagger \phi)$
    & any
    & any
    & 6
    \\
    $\mathbf{O}_{\phi \square 2}$
    & $(\scalar^\dagger T^a \scalar) D^2 (\phi^\dagger \sigma^a \phi)$
    & $T > 0$
    & any
    & 6
    \\
    $\mathbf{O}_{\phi D 1}$
    & $(\scalar^\dagger \DLR{\mu}{} \scalar) (\phi^\dagger \DLR{}{\mu} \phi)$
    & any
    & any
    & 6
    \\
    $\mathbf{O}_{\phi D 2}$
    & $(\scalar^\dagger \DLR{\mu}{a} \scalar) (\phi^\dagger \DLR{}{a\mu} \phi)$
    & $T > 0$
    & any
    & 6
    \\
    $\mathbf{O}_{\phi D 3}$
    & $(\scalarTilde^\dagger \DLR{\mu}{} \scalar) (\widetilde{\phi}^\dagger \DLR{}{\mu} \phi)$
    & $T \in \mathbb{Z} + 1/2$
    & $1/2$
    & 6
    \\
    $\mathbf{O}_{\phi D 4}$
    & $(\scalarTilde^\dagger \DLR{\mu}{a} \scalar) (\widetilde{\phi}^\dagger \DLR{a\mu}{} \phi)$
    & $T \in \mathbb{Z} + 1/2$
    & $1/2$
    & 6
    \\
    \midrule
    $\mathbf{O}_{B}$
    & $(\scalar^\dagger \scalar) (B_{\mu\nu} B^{\mu\nu})$
    & any
    & any
    & 6
    \\
    $\mathbf{O}_{\widetilde{B}}$
    & $(\scalar^\dagger \scalar) (B_{\mu\nu} \widetilde{B}^{\mu\nu})$
    & any
    & any
    & 6
    \\
    $\mathbf{O}_{W 1}$
    & $(\scalar^\dagger \scalar) (W_{\mu\nu}^a W^{a\mu\nu})$
    & any
    & any
    & 6
    \\
    $\mathbf{O}_{\widetilde{W} 1}$
    & $(\scalar^\dagger \scalar) (W_{\mu\nu}^a \widetilde{W}^{a\mu\nu})$
    & any
    & any
    & 6
    \\
    $\mathbf{O}_{W 2}$
    & $i\epsilon_{abc} (\scalar^\dagger T^a \scalar) (W^b_{\mu\nu} W^{c\mu\nu})$
    & $T > 1/2$
    & any
    & 6
    \\
    $\mathbf{O}_{\widetilde{W} 2}$
    & $i\epsilon_{abc} (\scalar^\dagger T^a \scalar) (W^b_{\mu\nu} \widetilde{W}^{c\mu\nu})$
    & $T > 1/2$
    & any
    & 6
    \\
    $\mathbf{O}_{BW}$
    & $(\scalar^\dagger T^a \scalar) (B_{\mu\nu} W^{a\mu\nu})$
    & $T > 0$
    & any
    & 6
    \\
    $\mathbf{O}_{B\widetilde{W}}$
    & $(\scalar^\dagger T^a \scalar) (B_{\mu\nu} \widetilde{W}^{a\mu\nu})$
    & $T > 0$
    & any
    & 6
    \\
    $\mathbf{O}_{G}$
    & $(\scalar^\dagger \scalar) (G_{\mu\nu}^A G^{A\mu\nu})$
    & any
    & any
    & 6
    \\
    $\mathbf{O}_{\widetilde{G}}$
    & $(\scalar^\dagger \scalar) (G_{\mu\nu}^A \widetilde{G}^{A\mu\nu})$
    & any
    & any
    & 6
    \\
    \bottomrule
  \end{tabular}
  \caption{Basis of operators of dimension $\leq 6$ with two scalar DM multiplets and SM bosons only.}
  \label{tab:scalar-basis-bosons}
\end{table}

\begin{table}[!ht]
\renewcommand{\arraystretch}{1.25}
  \centering
  \begin{tabular}{ccccc}
    \toprule
    Name
    & Operator
    & SU(2) irrep
    & Hypercharge
    & Dimension
    \\
    \midrule
    $\mathbf{O}_{l5}$
    & $(\scalarTilde^\dagger T^a \scalar) (\overline{\tilde{l}} \sigma^a l)$
    & $T \in \mathbb{Z} + 1/2$
    & $1/2$
    & 5
    \\
    \midrule
    $\mathbf{O}_{e}$
    & $(\scalar^\dagger i\DLR{\mu}{} \scalar) (\bar{e} \gamma^\mu e)$
    & any
    & any
    & 6
    \\
    $\mathbf{O}_{d}$
    & $(\scalar^\dagger i\DLR{\mu}{} \scalar) (\bar{d} \gamma^\mu d)$
    & any
    & any
    & 6
    \\
    $\mathbf{O}_{u}$
    & $(\scalar^\dagger i\DLR{\mu}{} \scalar) (\bar{u} \gamma^\mu u)$
    & any
    & any
    & 6
    \\
    $\mathbf{O}_{l 1}$
    & $(\scalar^\dagger i\DLR{\mu}{} \scalar) (\bar{l} \gamma^\mu l)$
    & any
    & any
    & 6
    \\
    $\mathbf{O}_{l 2}$
    & $(\scalar^\dagger i\DLR{\mu}{} \scalar) (\bar{l} \gamma^\mu \sigma^a l)$
    & $T > 0$
    & any
    & 6
    \\
    $\mathbf{O}_{q 1}$
    & $(\scalar^\dagger i\DLR{\mu}{} \scalar) (\bar{q} \gamma^\mu q)$
    & any
    & any
    & 6
    \\
    $\mathbf{O}_{q 2}$
    & $(\scalar^\dagger \DLR{\mu}{a} \scalar) (\bar{q} \gamma^\mu \sigma^a q)$
    & $T > 0$
    & any
    & 6
    \\
    $\mathbf{O}_{ud}$ 
    & $(\scalarTilde^\dagger \DLR{\mu}{} \scalar) (\bar{u} \gamma^\mu d)$
    & $T \in \mathbb{Z} + 1/2$
    & $1/2$
    & 6
    \\
    \midrule
    $\mathbf{O}_{e\phi 1}$
    & $(\scalar^\dagger \scalar) (\bar{l} \phi e)$
    & any
    & any
    & 6
    \\
    $\mathbf{O}_{e\phi 2}$
    & $(\scalar^\dagger T^a \scalar) (\bar{l} \sigma^a \phi e)$
    & $T > 0$
    & any
    & 6
    \\
    $\mathbf{O}_{e\phi 3}$
    & $(\scalarTilde^\dagger T^a \scalar) (\bar{l} \sigma^a \widetilde{\phi} e)$
    & any
    & $1/2$
    & 6
    \\
    $\mathbf{O}_{d\phi 1}$
    & $(\scalar^\dagger \scalar) (\bar{q} \phi d)$
    & any
    & any
    & 6
    \\
    $\mathbf{O}_{d\phi 2}$
    & $(\scalar^\dagger T^a \scalar) (\bar{q} \sigma^a \phi d)$
    & $T > 0$
    & any
    & 6
    \\
    $\mathbf{O}_{d\phi 3}$
    & $(\scalarTilde^\dagger T^a \scalar) (\bar{q} \sigma^a \widetilde{\phi} d)$
    & any
    & $1/2$
    & 6
    \\
    $\mathbf{O}_{u\phi 1}$
    & $(\scalar^\dagger \scalar) (\bar{q} \widetilde{\phi} u)$
    & any
    & any
    & 6
    \\
    $\mathbf{O}_{u\phi 2}$
    & $(\scalar^\dagger T^a \scalar) (\bar{q} \sigma^a \widetilde{\phi} u)$
    & $T > 0$
    & any
    & 6
    \\
    $\mathbf{O}_{u\phi 3}$
    & $(\scalarTilde^\dagger T^a \scalar) (\bar{q} \sigma^a \phi u)$
    & any
    & $1/2$
    & 6
    \\
    \bottomrule
  \end{tabular}
  \caption{Basis of operators of dimension $\leq 6$ with two scalar DM multiplets and at least one SM fermion.}
  \label{tab:scalar-basis-fermions}
\end{table}

\clearpage

\subsection{Operators for a fermion multiplet $\fermion$}

\begin{table}[!h]
  \centering
  \begin{tabular}{cccccc}
    \toprule
    Name
    & Operator
    & SU(2) irrep
    & Hypercharge
    & Dimension
    \\
    \midrule
    $\mathbf{O}_{\phi 1}$
    & $(\bar{\fermion}_L \fermion_R) (\phi^\dagger \phi)$
    & any & any & 5
    \\
    $\mathbf{O}_{\phi 2}$
    & $(\bar{\fermion}_L T^a \fermion_R) (\phi^\dagger \sigma^a \phi)$
    & $T > 0$ & any & 5
    \\
    $\mathbf{O}_{\phi 3}$
    & $(\bar{\fermion}_L T^a \widetilde{\fermion}_L) (\tilde{\phi}^\dagger \sigma^a \phi)$
    & $T \in \mathbb{Z} + 1/2$ & 1/2 & 5
    \\
    $\mathbf{O}_{\phi 4}$
    & $(\bar{\fermion}_R T^a \widetilde{\fermion}_R) (\tilde{\phi}^\dagger \sigma^a \phi)$
    & $T \in \mathbb{Z} + 1/2$ & 1/2 & 5
    \\
    $\mathbf{O}_{\phi 5}$
    & $(\bar{\fermion}_L \gamma^\mu \fermion_L) (\phi^\dagger \DLR{\mu}{} \phi)$
    & any & any & 6
    \\
    $\mathbf{O}_{\phi 6}$
    & $(\bar{\fermion}_R \gamma^\mu \fermion_R) (\phi^\dagger \DLR{\mu}{} \phi)$
    & any & any & 6
    \\
    $\mathbf{O}_{\phi 7}$
    & $(\bar{\fermion}_L T^a \gamma^\mu \fermion_L) (\phi^\dagger \DLR{\mu}{a} \phi)$
    & $T > 0$ & any & 6
    \\
    $\mathbf{O}_{\phi 8}$
    & $(\bar{\fermion}_R T^a \gamma^\mu \fermion_R) (\phi^\dagger \DLR{\mu}{a} \phi)$
    & $T > 0$ & any & 6
    \\
    $\mathbf{O}_{\phi 9}$
    & $(\bar{\fermion}_L \gamma^\mu \widetilde{\fermion}_R) (\tilde{\phi}^\dagger \DLR{\mu}{} \phi)$
    & $T > 0$ & 1/2 & 6
    \\
    \midrule
    $\mathbf{O}_B$
    & $\bar{\fermion}_L \sigma^{\mu\nu} \fermion_R B_{\mu\nu}$
    & any & any & 5
    \\
    $\mathbf{O}_W$
    & $\bar{\fermion}_L \sigma^{\mu\nu} T^a \fermion_R W^a_{\mu\nu}$
    & $T > 0$ & any & 5
    \\
    \midrule
    $\mathbf{O}_{e1}$
    & $(\bar{\fermion}_L \gamma^\mu \fermion_L) (\bar{e} \gamma_\mu e)$
    & any & any & 6
    \\
    $\mathbf{O}_{e2}$
    & $(\bar{\fermion}_R \gamma^\mu \fermion_R) (\bar{e} \gamma_\mu e)$
    & any & any & 6
    \\
    $\mathbf{O}_{d1}$
    & $(\bar{\fermion}_L \gamma^\mu \fermion_L) (\bar{d} \gamma_\mu d)$
    & any & any & 6
    \\
    $\mathbf{O}_{d2}$
    & $(\bar{\fermion}_R \gamma^\mu \fermion_R) (\bar{d} \gamma_\mu d)$
    & any & any & 6
    \\
    $\mathbf{O}_{u1}$
    & $(\bar{\fermion}_L \gamma^\mu \fermion_L) (\bar{u} \gamma_\mu u)$
    & any & any & 6
    \\
    $\mathbf{O}_{u2}$
    & $(\bar{\fermion}_R \gamma^\mu \fermion_R) (\bar{u} \gamma_\mu u)$
    & any & any & 6
    \\
    $\mathbf{O}_{ud}$
    & $(\bar{\fermion}_L^c \gamma^\mu \fermion_R) (\bar{u} \gamma_\mu d)$
    & any & any & 6
    \\
    $\mathbf{O}_{l1}$
    & $(\bar{\fermion}_L \gamma^\mu \fermion_L) (\bar{l} \gamma_\mu l)$
    & any & any & 6
    \\
    $\mathbf{O}_{l2}$
    & $(\bar{\fermion}_R \gamma^\mu \fermion_R) (\bar{l} \gamma_\mu l)$
    & any & any & 6
    \\
    $\mathbf{O}_{l3}$
    & $(\bar{\fermion}_L T^a \gamma^\mu \fermion_L) (\bar{l} \sigma^a \gamma_\mu l)$
    & $T > 0$ & any & 6
    \\
    $\mathbf{O}_{l4}$
    & $(\bar{\fermion}_R T^a \gamma^\mu \fermion_R) (\bar{l} \sigma^a \gamma_\mu l)$
    & $T > 0$ & any & 6
    \\
    $\mathbf{O}_{l5}$
    & $(\bar{\fermion}_L \widetilde{\fermion}_L) (\bar{l} \tilde{l})$
    & $T \in \mathbb{Z} + 1/2$ & 1/2 & 6
    \\
    $\mathbf{O}_{l6}$
    & $(\bar{\fermion}_L T^a \widetilde{\fermion}_L) (\bar{l} \sigma^a \tilde{l})$
    & $T \in \mathbb{Z} + 1/2$ & 1/2 & 6
    \\
    $\mathbf{O}_{l7}$
    & $(\bar{\fermion}_R \widetilde{\fermion}_R) (\bar{l} \tilde{l})$
    & $T \in \mathbb{Z} + 1/2$ & 1/2 & 6
    \\
    $\mathbf{O}_{q1}$
    & $(\bar{\fermion}_L \gamma^\mu \fermion_L) (\bar{q} \gamma_\mu q)$
    & any & any & 6
    \\
    $\mathbf{O}_{q2}$
    & $(\bar{\fermion}_R \gamma^\mu \fermion_R) (\bar{q} \gamma_\mu q)$
    & any & any & 6
    \\
    $\mathbf{O}_{q3}$
    & $(\bar{\fermion}_L T^a \gamma^\mu \fermion_L) (\bar{q} \sigma^a \gamma_\mu q)$
    & $T > 0$ & any & 6
    \\
    $\mathbf{O}_{q4}$
    & $(\bar{\fermion}_R T^a \gamma^\mu \fermion_R) (\bar{q} \sigma^a \gamma_\mu q)$
    & $T > 0$ & any & 6
    \\
    \bottomrule
  \end{tabular}
  \caption{Basis of operators of dimension $\leq 6$ with two fermion DM multiplets.}
  \label{tab:fermion-basis}
\end{table}

\clearpage

\subsection{Operators for a vector multiplet $\vector$}

\begin{table}[!h]
\renewcommand{\arraystretch}{1.25}
  \centering
  \begin{tabular}{ccccc}
    \toprule
    Name
    & Operator
    & SU(2) irrep
    & Hypercharge & Dimension
    \\
    \midrule
    $\mathbf{O}_{B 1}$
    & $(\vector^\dagger_\mu \vector_\nu) B^{\mu\nu}$
    & any
    & any & 4
    \\
    $\mathbf{O}_{\widetilde{B} 1}$
    & $(\vector^\dagger_\mu \vector_\nu) \widetilde{B}^{\mu\nu}$
    & any
    & any & 4
    \\
    $\mathbf{O}_{B 2}$
    & $(\vector^\dagger_\mu \vector^\nu) B^{\mu\vector} B_{\vector\nu}$
    & any
    & any & 6
    \\
    $\mathbf{O}_{\widetilde{B} 2}$
    & $(\vector^\dagger_\mu \vector^\nu) \widetilde{B}^{\mu\vector} \widetilde{B}_{\vector\nu}$
    & any
    & any & 6
    \\
    $\mathbf{O}_{B\widetilde{B}}$
    & $(\vector^\dagger_\mu \vector^\nu) B^{\mu\vector} \widetilde{B}_{\vector\nu}$
    & any
    & any & 6
    \\
    $\mathbf{O}_{W 1}$
    & $(\vector^\dagger_\mu \sigma^a \vector_\nu) W^{a\mu\nu}$
    & $T > 0$
    & any & 4
    \\
    $\mathbf{O}_{\widetilde{W} 1}$
    & $(\vector^\dagger_\mu \sigma^a \vector_\nu) \widetilde{W}^{a\mu\nu}$
    & $T > 0$
    & any & 4
    \\
    $\mathbf{O}_{W 2}$
    & $(\vector^\dagger_\mu \vector^\nu) W^{a \mu\vector} W^a_{\vector\nu}$
    & any
    & any & 6
    \\
    $\mathbf{O}_{W 3}$
    & $\epsilon_{abc} (\vector^\dagger_\mu \sigma^a \vector^\nu) W^{b \mu\vector} W^c_{\vector\nu}$
    & $T > 0$
    & any & 6
    \\
    $\mathbf{O}_{W 4}$
    & $C_{Iab} (\vector^\dagger_\mu Q^I \vector^\nu) W^{a \mu\vector} W^b_{\vector\nu}$
    & $T > 1/2$
    & any & 6
    \\
    $\mathbf{O}_{\widetilde{W} 2}$
    & $(\vector^\dagger_\mu \vector^\nu) \widetilde{W}^{a \mu\vector} \widetilde{W}^a_{\vector\nu}$
    & any
    & any & 6
    \\
    $\mathbf{O}_{\widetilde{W} 3}$
    & $\epsilon_{abc} (\vector^\dagger_\mu \sigma^a \vector^\nu) \widetilde{W}^{b \mu\vector} \widetilde{W}^c_{\vector\nu}$
    & $T > 0$
    & any & 6
    \\
    $\mathbf{O}_{\widetilde{W} 4}$
    & $C_{Iab} (\vector^\dagger_\mu Q^I \vector^\nu) \widetilde{W}^{a \mu\vector} \widetilde{W}^b_{\vector\nu}$
    & $T > 1/2$
    & any & 6
    \\
    $\mathbf{O}_{W\widetilde{W} 1}$
    & $(\vector^\dagger_\mu \vector^\nu) W^{a \mu\vector} \widetilde{W}^a_{\vector\nu}$
    & any
    & any & 6
    \\
    $\mathbf{O}_{W\widetilde{W} 2}$
    & $\epsilon_{abc} (\vector^\dagger_\mu \sigma^a \vector^\nu) W^{b \mu\vector} \widetilde{W}^c_{\vector\nu}$
    & $T > 0$
    & any & 6
    \\
    $\mathbf{O}_{W\widetilde{W} 3}$
    & $C_{Iab} (\vector^\dagger_\mu Q^I \vector^\nu) W^{a \mu\vector} \widetilde{W}^b_{\vector\nu}$
    & $T > 1/2$
    & any & 6
    \\
    $\mathbf{O}_{BW 1}$
    & $(\vector^\dagger_\mu \sigma^a \vector^\mu) B_{\nu\vector} W^{a \nu\vector}$
    & $T > 0$
    & any & 6
    \\
    $\mathbf{O}_{BW 2}$
    & $(\vector^\dagger_\mu \sigma^a \vector^\nu) B^{\mu\vector} W_{a \nu\vector}$
    & $T > 0$
    & any & 6
    \\
    $\mathbf{O}_{\widetilde{B}\widetilde{W} 1}$
    & $(\vector^\dagger_\mu \sigma^a \vector^\mu) \widetilde{B}_{\nu\vector} \widetilde{W}^{a \nu\vector}$
    & $T > 0$
    & any & 6
    \\
    $\mathbf{O}_{\widetilde{B}\widetilde{W} 2}$
    & $(\vector^\dagger_\mu \sigma^a \vector^\nu) \widetilde{B}^{\mu\vector} \widetilde{W}_{a \nu\vector}$
    & $T > 0$
    & any & 6
    \\
    $\mathbf{O}_{B\widetilde{W} 1}$
    & $(\vector^\dagger_\mu \sigma^a \vector^\mu) B_{\nu\vector} \widetilde{W}^{a \nu\vector}$
    & $T > 0$
    & any & 6
    \\
    $\mathbf{O}_{B\widetilde{W} 2}$
    & $(\vector^\dagger_\mu \sigma^a \vector^\nu) B^{\mu\vector} \widetilde{W}_{a \nu\vector}$
    & $T > 0$
    & any & 6
    \\
    $\mathbf{O}_{GG}$
    & $(\vector^\dagger_\mu \vector^\mu) G^{A \nu\vector} \widetilde{G}^A_{\nu\vector}$
    & any
    & any & 6
    \\
    $\mathbf{O}_{G\widetilde{G}}$
    & $(\vector^\dagger_\mu \vector^\mu) G^{A \nu\vector} \widetilde{G}^A_{\nu\vector}$
    & any
    & any & 6
    \\
    $\mathbf{O}_{\widetilde{G}\widetilde{G}}$
    & $(\vector^\dagger_\mu \vector^\mu) \widetilde{G}^{A \nu\vector} \widetilde{G}^A_{\nu\vector}$
    & any
    & any & 6
    \\
    \bottomrule
  \end{tabular}
  \caption{Basis of operators of dimension $\leq 6$ with two vector DM multiplets and SM field-strength tensors only. $Q^I$ is the unique set of square matrices acting on the SU(2) part of $\vector$ which transform covariantly as a quadruplet. $C_{Iab}$ represents the quadruplet-triplet-triplet Clebsh-Gordan coefficients.}
  \label{tab:vector-basis-tensors}
\end{table}

\begin{table}
  \centering
  \begin{tabular}{ccccc}
    \toprule
    Name
    & Operator
    & SU(2) irrep
    & Hypercharge & Dimension
    \\
    \midrule
    $\mathbf{O}_{e\phi 1}$
    & $(\vector^\dagger_\mu \vector^\mu) (\bar{l} \phi e)$
    & any
    & any & 6
    \\
    $\mathbf{O}_{e\phi 2}$
    & $(\vector^\dagger_\mu \vector_\nu) (\bar{l} \sigma^{\mu\nu} \phi e)$
    & any
    & any & 6
    \\
    $\mathbf{O}_{e\phi 3}$
    & $(\vector^\dagger_\mu T^a \vector^\mu) (\bar{l} \sigma^a \phi e)$
    & $T > 0$
    & any & 6
    \\
    $\mathbf{O}_{e\phi 4}$
    & $(\vector^\dagger_\mu T^a \vector_\nu) (\bar{l} \sigma^{\mu\nu} \sigma^a \phi e)$
    & $T > 0$
    & any & 6
    \\
    $\mathbf{O}_{e\phi 5}$
    & $(\vector^\dagger_\mu \widetilde{\vector}^\mu) (\bar{e} \widetilde{\phi}^\dagger l)$
    & any
    & 1/2 & 6
    \\
    $\mathbf{O}_{e\phi 6}$
    & $(\vector^\dagger_\mu T^a \widetilde{\vector}^\mu) (\bar{e} \widetilde{\phi}^\dagger \sigma^a l)$
    & $T > 0$
    & 1/2 & 6
    \\
    $\mathbf{O}_{d\phi 1}$
    & $(\vector^\dagger_\mu \vector^\mu) (\bar{q} \phi d)$
    & any
    & any & 6
    \\
    $\mathbf{O}_{d\phi 2}$
    & $(\vector^\dagger_\mu \vector_\nu) (\bar{q} \sigma^{\mu\nu} \phi d)$
    & any
    & any & 6
    \\
    $\mathbf{O}_{d\phi 3}$
    & $(\vector^\dagger_\mu T^a \vector^\mu) (\bar{q} \sigma^a \phi d)$
    & $T > 0$
    & any & 6
    \\
    $\mathbf{O}_{d\phi 4}$
    & $(\vector^\dagger_\mu T^a \vector_\nu) (\bar{q} \sigma^{\mu\nu} \sigma^a \phi d)$
    & $T > 0$
    & any & 6
    \\
    $\mathbf{O}_{d\phi 5}$
    & $(\vector^\dagger_\mu \widetilde{\vector}^\mu) (\bar{d} \widetilde{\phi}^\dagger q)$
    & any
    & 1/2 & 6
    \\
    $\mathbf{O}_{d\phi 6}$
    & $(\vector^\dagger_\mu T^a \widetilde{\vector}^\mu) (\bar{d} \widetilde{\phi}^\dagger \sigma^a q)$
    & $T > 0$
    & 1/2 & 6
    \\
    $\mathbf{O}_{u\phi 1}$
    & $(\vector^\dagger_\mu \vector^\mu) (\bar{q} \widetilde{\phi} u)$
    & any
    & any & 6
    \\
    $\mathbf{O}_{u\phi 2}$
    & $(\vector^\dagger_\mu \vector_\nu) (\bar{q} \sigma^{\mu\nu} \widetilde{\phi} u)$
    & any
    & any & 6
    \\
    $\mathbf{O}_{u\phi 3}$
    & $(\vector^\dagger_\mu T^a \vector^\mu) (\bar{q} \sigma^a \widetilde{\phi} u)$
    & $T > 0$
    & any & 6
    \\
    $\mathbf{O}_{u\phi 4}$
    & $(\vector^\dagger_\mu T^a \vector_\nu) (\bar{q} \sigma^{\mu\nu} \sigma^a \widetilde{\phi} u)$
    & $T > 0$
    & any & 6
    \\
    $\mathbf{O}_{u\phi 5}$
    & $(\vector^\dagger_\mu \widetilde{\vector}^\mu) (\bar{q} \phi u)$
    & any
    & 1/2 & 6
    \\
    $\mathbf{O}_{u\phi 6}$
    & $(\vector^\dagger_\mu T^a \widetilde{\vector}^\mu) (\bar{q} \sigma^a \phi u)$
    & $T > 0$
    & 1/2 & 6
    \\
    \midrule
    $\mathbf{O}_{e 1}$
    & $(\vector^\dagger_\mu D_\nu \vector^\mu) (\bar{e} \gamma^\nu e)$
    & any
    & any & 6
    \\
    $\mathbf{O}_{e 2}$
    & $(\vector^\dagger_\mu D_\nu \vector^\nu) (\bar{e} \gamma^\mu e)$
    & any
    & any & 6
    \\
    $\mathbf{O}_{d 1}$
    & $(\vector^\dagger_\mu D_\nu \vector^\mu) (\bar{d} \gamma^\nu d)$
    & any
    & any & 6
    \\
    $\mathbf{O}_{d 2}$
    & $(\vector^\dagger_\mu D_\nu \vector^\nu) (\bar{d} \gamma^\mu d)$
    & any
    & any & 6
    \\
    $\mathbf{O}_{u 1}$
    & $(\vector^\dagger_\mu D_\nu \vector^\mu) (\bar{u} \gamma^\nu u)$
    & any
    & any & 6
    \\
    $\mathbf{O}_{u 2}$
    & $(\vector^\dagger_\mu D_\nu \vector^\nu) (\bar{u} \gamma^\mu u)$
    & any
    & any & 6
    \\
    $\mathbf{O}_{ud}$
    & $(\vector^\dagger_\mu D_\nu \tilde{\vector}^\mu) (\bar{u} \gamma^\nu d)$
    & any
    & 1/2 & 6
    \\
    $\mathbf{O}_{l1}$
    & $(\vector^\dagger_\mu T^a \tilde{\vector}^\mu) (\bar{l} \sigma^a \tilde{l})$
    & $T \in \mathbb{Z} + 1/2$
    & 1/2 & 6
    \\
    $\mathbf{O}_{l2}$
    & $(\vector^\dagger_\mu \tilde{\vector}_\nu) (\bar{l} \sigma^{\mu\nu} \tilde{l})$
    & $T \in \mathbb{Z} + 1/2$
    & 1/2 & 6
    \\
    $\mathbf{O}_{l 3}$
    & $(\vector^\dagger_\mu D_\nu \vector^\mu) (\bar{l} \gamma^\nu l)$
    & any
    & any & 6
    \\
    $\mathbf{O}_{l 4}$
    & $(\vector^\dagger_\mu D_\nu \vector^\nu) (\bar{l} \gamma^\mu l)$
    & any
    & any & 6
    \\
    $\mathbf{O}_{l 5}$
    & $(\vector^\dagger_\mu T^a D_\nu \vector^\mu) (\bar{l} \gamma^\nu \sigma^a l)$
    & $T > 0$
    & any & 6
    \\
    $\mathbf{O}_{l 6}$
    & $(\vector^\dagger_\mu T^a D_\nu \vector^\nu) (\bar{l} \gamma^\mu \sigma^a l)$
    & $T > 0$
    & any & 6
    \\
    $\mathbf{O}_{q 1}$
    & $(\vector^\dagger_\mu D_\nu \vector^\mu) (\bar{q} \gamma^\nu q)$
    & any
    & any & 6
    \\
    $\mathbf{O}_{q 2}$
    & $(\vector^\dagger_\mu D_\nu \vector^\nu) (\bar{q} \gamma^\mu q)$
    & any
    & any & 6
    \\
    $\mathbf{O}_{q 3}$
    & $(\vector^\dagger_\mu T^a D_\nu \vector^\mu) (\bar{q} \gamma^\nu \sigma^a q)$
    & $T > 0$
    & any & 6
    \\
    $\mathbf{O}_{q 4}$
    & $(\vector^\dagger_\mu T^a D_\nu \vector^\nu) (\bar{q} \gamma^\mu \sigma^a q)$
    & $T > 0$
    & any & 6
    \\
    \bottomrule
  \end{tabular}
  \caption{Basis of operators of dimension $\leq 6$ with two vector DM multiplets and at least two SM fermions.}
  \label{tab:vector-basis-fermions}
\end{table}

\begin{table}
  \centering
  \begin{tabular}{ccccc}
    \toprule
    Name
    & Operator
    & SU(2) irrep
    & Hypercharge
    & Dimension
    \\
    \midrule
    $\mathbf{O}_{\phi 1}$
    & $(\vector^\dagger_\mu \vector^\mu) (\phi^\dagger \phi)$
    & any
    & any & 4
    \\
    $\mathbf{O}_{\phi 2}$
    & $(\vector^\dagger_\mu T^a \vector^\mu) (\phi^\dagger \sigma^a \phi)$
    & $T > 0$
    & any & 4
    \\
    $\mathbf{O}_{\phi 3}$
    & $(\vector^\dagger_\mu T^a \widetilde{\vector}^\mu) (\widetilde{\phi}^\dagger \sigma^a \phi)$
    & $T \in \mathbb{Z} + 1/2$
    & 1/2 & 4
    \\
    $\mathbf{O}_{\phi 4}$
    & $(\vector^\dagger_\mu \vector^\mu) (\phi^\dagger \phi)^2$
    & any
    & any & 6
    \\
    $\mathbf{O}_{\phi 5}$
    & $(\vector^\dagger_\mu T^a \vector^\mu) (\phi^\dagger \sigma^a \phi) (\phi^\dagger \phi)$
    & $T > 0$
    & any & 6
    \\
    $\mathbf{O}_{\phi 6}$
    & $\epsilon_{abc} (\vector^\dagger_\mu T^a \vector^\mu)
      (\phi^\dagger \sigma^b \widetilde{\phi})
      (\widetilde{\phi}^\dagger \sigma^c \phi)$
    & $T > 1/2$
    & any & 6
    \\
    $\mathbf{O}_{\phi 7}$
    & $(\widetilde{\vector}^\dagger_\mu T^a \vector^\mu) (\phi^\dagger \sigma^a \widetilde{\phi}) (\phi^\dagger \phi)$
    & $T > 0$
    & any & 6
    \\
    $\mathbf{O}_{\phi \square 1}$
    & $(\vector^\dagger_\mu \vector^\mu) \square (\phi^\dagger \phi)$
    & any
    & any & 6
    \\
    $\mathbf{O}_{\phi \square 2}$
    & $(\vector^\dagger_\mu T^a \vector^\mu) D^2 (\phi^\dagger \sigma^a \phi)$
    & $T > 0$
    & any & 6
    \\
    $\mathbf{O}_{\phi \square 3}$
    & $(\vector^\dagger_\mu T^a \widetilde{\vector}^\mu) D^2 (\widetilde{\phi}^\dagger \sigma^a \phi)$
    & $T \in \mathbb{Z} + 1/2$
    & 1/2 & 6
    \\
    $\mathbf{O}_{\phi D 1}$
    & $(\vector^\dagger_\mu \DLR{}{\nu} \vector^\mu) (\phi^\dagger \DLR{\nu}{} \phi)$
    & any
    & any & 6
    \\
    $\mathbf{O}_{\phi D 2}$
    & $(\vector^\dagger_\mu \DLR{}{\nu} \vector^\nu) (\phi^\dagger \DLR{\nu}{} \phi)$
    & any
    & any & 6
    \\
    $\mathbf{O}_{\phi D 3}$
    & $(\vector^\dagger_\mu \DLR{}{a\nu} \vector^\mu) (\phi^\dagger \DLR{\nu}{a} \phi)$
    & $T > 0$
    & any & 6
    \\
    $\mathbf{O}_{\phi D 4}$
    & $(\vector^\dagger_\mu \DLR{}{a\nu} \vector^\nu) (\phi^\dagger \DLR{\nu}{a} \phi)$
    & $T > 0$
    & any & 6
    \\
    $\mathbf{O}_{\phi D 5}$
    & $(\vector^\dagger_\mu \DLR{}{a\nu} \widetilde{\vector}^\mu) (\widetilde{\phi}^\dagger \DLR{\nu}{a} \phi)$
    & $T \in \mathbb{Z} + 1/2$
    & 1/2 & 6
    \\
    $\mathbf{O}_{\phi D 6}$
    & $(\vector^\dagger_\mu \DLR{}{a\nu} \widetilde{\vector}^\nu) (\widetilde{\phi}^\dagger \DLR{\nu}{a} \phi)$
    & $T \in \mathbb{Z} + 1/2$
    & 1/2 & 6
    \\
    \midrule
    $\mathbf{O}_{\phi B 1}$
    & $(\vector^\dagger_\mu \vector_\nu) B^{\mu\nu} (\phi^\dagger \phi)$
    & any
    & any & 6
    \\
    $\mathbf{O}_{\phi B 2}$
    & $(\vector^\dagger_\mu T^a \vector_\nu) B^{\mu\nu} (\phi^\dagger \sigma^a \phi)$
    & $T > 0$
    & any & 6
    \\
    $\mathbf{O}_{\phi B 3}$
    & $(\vector^\dagger_\mu T^a \widetilde{\vector}_\nu) B^{\mu\nu} (\widetilde{\phi}^\dagger \sigma^a \phi)$
    & $T \in \mathbb{Z} + 1/2$
    & 1/2 & 6
    \\
    $\mathbf{O}_{\phi \widetilde{B} 1}$
    & $(\vector^\dagger_\mu \vector_\nu) \widetilde{B}^{\mu\nu} (\phi^\dagger \phi)$
    & any
    & any & 6
    \\
    $\mathbf{O}_{\phi \widetilde{B} 2}$
    & $(\vector^\dagger_\mu T^a \vector_\nu) \widetilde{B}^{\mu\nu} (\phi^\dagger \sigma^a \phi)$
    & $T > 0$
    & any & 6
    \\
    $\mathbf{O}_{\phi \widetilde{B} 3}$
    & $(\vector^\dagger_\mu T^a \widetilde{\vector}_\nu) \widetilde{B}^{\mu\nu} (\widetilde{\phi}^\dagger \sigma^a \phi)$
    & $T \in \mathbb{Z} + 1/2$
    & 1/2 & 6
    \\
    $\mathbf{O}_{\phi W 1}$
    & $(\vector^\dagger_\mu \vector_\nu) W^{a \mu\nu} (\phi^\dagger \sigma^a \phi)$
    & any
    & any & 6
    \\
    $\mathbf{O}_{\phi W 2}$
    & $(\vector^\dagger_\mu T^a \vector_\nu) W^{a \mu\nu} (\phi^\dagger \phi)$
    & $T > 0$
    & any & 6
    \\
    $\mathbf{O}_{\phi W 3}$
    & $\epsilon_{abc} (\vector^\dagger_\mu T^a \vector_\nu) W^{b \mu\nu} (\phi^\dagger \sigma^c \phi)$
    & $T > 0$
    & any & 6
    \\
    $\mathbf{O}_{\phi W 4}$
    & $C_{Iab} (\vector^\dagger_\mu Q^I \vector_\nu) W^{a \mu\nu} (\phi^\dagger \sigma^b \phi)$
    & $T > 1/2$
    & any & 6
    \\
    $\mathbf{O}_{\phi W 5}$
    & $(\vector^\dagger_\mu \widetilde{\vector}_\nu) W^{a \mu\nu} (\widetilde{\phi}^\dagger \sigma^a \phi)$
    & any
    & 1/2 & 6
    \\
    $\mathbf{O}_{\phi W 6}$
    & $(\vector^\dagger_\mu T^a \widetilde{\vector}_\nu) W^{a \mu\nu} (\widetilde{\phi}^\dagger \phi)$
    & $1/2 < T \in \mathbb{Z} + 1/2$
    & 1/2 & 6
    \\
    $\mathbf{O}_{\phi \widetilde{W} 1}$
    & $(\vector^\dagger_\mu \vector_\nu) \widetilde{W}^{a \mu\nu} (\phi^\dagger \sigma^a \phi)$
    & any
    & any & 6
    \\
    $\mathbf{O}_{\phi \widetilde{W} 2}$
    & $(\vector^\dagger_\mu T^a \vector_\nu) \widetilde{W}^{a \mu\nu} (\phi^\dagger \phi)$
    & $T > 1/2$
    & any & 6
    \\
    $\mathbf{O}_{\phi \widetilde{W} 3}$
    & $\epsilon_{abc} (\vector^\dagger_\mu T^a \vector_\nu) \widetilde{W}^{b \mu\nu} (\phi^\dagger \sigma^c \phi)$
    & $T > 1/2$
    & any & 6
    \\
    $\mathbf{O}_{\phi \widetilde{W} 4}$
    & $C_{Iab} (\vector^\dagger_\mu Q^I \vector_\nu) \widetilde{W}^{a \mu\nu} (\phi^\dagger \sigma^b \phi)$
    & $T > 1$
    & any & 6
    \\
    $\mathbf{O}_{\phi \widetilde{W} 5}$
    & $(\vector^\dagger_\mu \widetilde{\vector}_\nu) \widetilde{W}^{a \mu\nu} (\widetilde{\phi}^\dagger \sigma^a \phi)$
    & any
    & 1/2 & 6
    \\
    $\mathbf{O}_{\phi \widetilde{W} 6}$
    & $(\vector^\dagger_\mu T^a \widetilde{\vector}_\nu) \widetilde{W}^{a \mu\nu} (\widetilde{\phi}^\dagger \phi)$
    & $T > 0$
    & 1/2 & 6
    \\
    \bottomrule
      \end{tabular}
  \caption{Basis of operators of dimension $\leq 6$ with two vector DM multiplets, at least one Higgs boson, and no SM fermions. $Q^I$ is the unique set of square matrices acting on the SU(2) part of $\vector$ which transform covariantly as a quadruplet. $C_{Iab}$ represents the quadruplet-triplet-triplet Clebsh-Gordan coefficients.}
  \label{tab:vector-basis-higgs}
\end{table}

\clearpage

\section{Basis of $\scalar^\dagger \scalar \, \phi^\dagger \phi D^2$}
\label{app:operators}

We prove here that the linear space of operators with the field content $\scalar^\dagger \scalar \, \phi^\dagger \phi D^2$ is parametrized by 4 real coefficients if $T > 0$, or 2 real coefficients if $T = 0$.
We first notice that there are two independent $\SU(2)$ structures for $T > 0$, without derivatives, given by $\mathcal{O}_{\phi 1} = (\scalar^\dagger \scalar) (\phi^\dagger \phi)$ and $\mathcal{O}_{\phi 2} = (\scalar^\dagger T^a \scalar) (\phi^\dagger \sigma^a \phi)$. For each of them, we will count the number of independent ways of introducing two derivatives in the operator. First, using the Leibniz rule for covariant derivatives, we can assume that each derivative is applied to one field only. Second, the application of two derivatives in the combination $D^2 = D_\mu D^\mu$ to any scalar field produces an operator that can be eliminated using equations of motion. This leaves $6$ possibilities:
\begin{gather}
    \mathbf{Q}_1 = (D_\mu \scalar^\dagger D^\mu \scalar) (\phi^\dagger \phi),
    \quad
    \mathbf{Q}_2 = (D_\mu \scalar^\dagger \scalar) (D^\mu \phi^\dagger \phi),
    \quad
    \mathbf{Q}_3 = (D_\mu \scalar^\dagger \scalar) (\phi^\dagger D^\mu \phi),
    \\
    \mathbf{Q}_4 = (\scalar^\dagger D_\mu \scalar) (D^\mu\phi^\dagger \phi),
    \quad
    \mathbf{Q}_5 = (\scalar^\dagger D_\mu \scalar) (\phi^\dagger D^\mu\phi),
    \quad
    \mathbf{Q}_6 = (\scalar^\dagger \scalar) (D_\mu\phi^\dagger D^\mu \phi),
\end{gather}
for the $\mathbf{O}_{\phi 1}$ structure and 6 other similar possibilities for the $\mathbf{O}_{\phi 2}$ one. Two of these possibilities are the complex conjugates of the others ($\mathbf{Q}_2 = \mathbf{Q}_5^\dagger$, $\mathbf{Q}_3 = \mathbf{Q}_4^\dagger$), but we include them explicitly in order to count real degrees of freedom. It remains to eliminate integration by parts redundancy. In order to do so, we define the following operators with one derivative:
\begin{gather}
    \mathbf{V}_{1\mu} = (D_\mu \scalar^\dagger \scalar) (\phi^\dagger \phi),
    \quad
    \mathbf{V}_{2\mu} = (\scalar^\dagger D_\mu \scalar) (\phi^\dagger \phi),
    \\
    \mathbf{V}_{3\mu} = (\scalar^\dagger \scalar) (D_\mu\phi^\dagger \phi),
    \quad
    \mathbf{V}_{4\mu} = (\scalar^\dagger \scalar) (\phi^\dagger D_\mu\phi).
\end{gather}
They generate $4$ linear relations $D^\mu \mathbf{V}_{i\mu} = 0$, $i = 1, \ldots, 4$, between $\mathbf{Q}_j$ operators, which are
\begin{equation}
    \begin{pmatrix}
        1 & 1 & 1 & 0 & 0 & 0 \\
        1 & 0 & 0 & 1 & 1 & 0 \\
        0 & 1 & 0 & 1 & 0 & 1 \\
        0 & 0 & 1 & 0 & 1 & 1
    \end{pmatrix}
    \begin{pmatrix}
        \mathbf{Q}_1 \\
        \mathbf{Q}_2 \\
        \mathbf{Q}_3 \\
        \mathbf{Q}_4 \\
        \mathbf{Q}_5 \\
        \mathbf{Q}_6
    \end{pmatrix}
    =
    \begin{pmatrix}
        0 \\ 0 \\ 0 \\ 0
    \end{pmatrix}.
\end{equation}
Since the rank of the matrix on the left-hand side is 4, the 4 relations are independent, leaving 2 independent hermitian operators with 2 derivatives, for each $SU(2)$ structure.

\bibliographystyle{JHEP}
\bibliography{references}

\end{document}